\documentclass[apl,twocolumn,showpacs,superscriptaddress,letterpaper]{revtex4-1}
\usepackage{graphicx,amsmath,amsfonts}
\usepackage{tabularx}
\usepackage{bm}
\usepackage{mathrsfs}
\usepackage{appendix}
\usepackage{bbm}
\usepackage{color}
\def \uL {{\underline L}}

\def \uP {{\underline P}}
\def \uQ {{\underline Q}}
\def \uPp {{\underline P}^{\prime }}
\def \uQp {{\underline Q}^{\prime }}

\def \vk {{\bf k}}

\def \vS {{\bf S}}

\def \mb {\mu_{\rm B}}

\def \vz {{\bf z}}

\def \vr {{\bf r}}

\def \vG {{\bf G}}
\def \xik {\xi_{\vk }}

\def \lk {{\hat{l}_{z\vk }}}

\def \kB {k_{\rm B}}

\begin{document}

\title{Magnon Orbital Angular Momentum of Ferromagnetic Honeycomb and Zig-Zag Lattices}

\author{Randy S. Fishman}
\email{corresponding author:  fishmanrs@ornl.gov
\newline
This manuscript has been authored in part by UT-Battelle, LLC, under contract DE-AC05-00OR22725 with the US Department of Energy (DOE). The US government retains and the publisher, by accepting the article for publication, acknowledges that the US government retains a nonexclusive, paid-up, irrevocable, worldwide license to publish or reproduce the published form of this manuscript, or allow others to do so, for US government purposes. DOE will provide public access to these results of federally sponsored research in accordance with the DOE Public Access Plan (http://energy.gov/downloads/doe-public-access-plan)}

\affiliation{Materials Science and Technology Division, Oak Ridge National Laboratory, Oak Ridge, Tennessee 37831, USA}
\author{Tom Berlijn}
\affiliation{Materials Science and Technology Division, Oak Ridge National Laboratory, Oak Ridge, Tennessee 37831, USA}
\author{Jack Villanova}
\affiliation{Center for Nanophase Materials Sciences, Oak Ridge National Laboratory, Oak Ridge, Tennessee 37831, USA}
\author{Lucas Lindsay}
\affiliation{Materials Science and Technology Division, Oak Ridge National Laboratory, Oak Ridge, Tennessee 37831, USA}

\date{\today}

\begin{abstract}  By expanding the gauge $\lambda_n(\vk )$ for magnon band $n$ in harmonics of momentum $\vk =(k,\phi )$,
we demonstrate that the only observable component of the magnon orbital angular momentum $O_n(\vk )$ is its angular average 
over all angles $\phi$, denoted by $F_n(k)$.  Although $F_n(k)$ vanishes for antiferromagnetic honeycomb and zig-zag ($0< J_1 < J_2$) lattices, it
is nonzero for the ferromagnetic (FM) versions of those lattices in the presence of
Dzyalloshinskii-Moriya (DM) interactions.
For a FM zig-zag model with equal exchange interactions $J_{1x}$ and $J_{1y}$ along the $x$ and $y$ axis, the magnon bands are degenerate
along the boundaries of the Brillouin zone with $k_x-k_y =\pm \pi/a$ and the Chern numbers $C_n$ are not well defined.  However, a revised model with 
$J_{1y}\ne J_{1x}$ lifts those degeneracy and produces well-defined Chern numbers of $C_n=\pm 1$ for the two magnon bands.  
When $J_{1y}=J_{1x}$, the thermal conductivity $\kappa^{xy}(T)$ of
the FM zig-zag lattice is largest for $J_2/J_1>6$ but is still about four times smaller than that of the FM honeycomb lattice
at high temperatures.  Due to the removal of band degeneracies, $\kappa^{xy}(T)$ is slightly enhanced when $J_{1y}\ne J_{1x}$.
		\end{abstract}


\maketitle

\section{Introduction}

The past 13 years have seen remarkable advances in the field of  ``magnonics" \cite{Wang21, Chumak22, Sheka22}, which focuses on 
the quanta of spin excitations known as magnons.  One of the main goals of magnonics is the storage and processing of information. 
Because they can travel over centimeter distances without
incurring any costs in Joule heating \cite{Buttner2000}, magnons offer many advantages over electrons in the next generation of
technological devices.  Due to their much lower velocities, magnons are also better suited than electrons 
to creating small devices.
In quick succession, experimentalists have discovered that magnons can produce the 
thermal Hall  \cite{Onose2010, Ideue12, Hirschberger15a, Hirschberger15b, Murakami17, Neumann22} and
Seebeck \cite{Uchida10, Wu16} effects.

Almost all previous theoretical work in magnonics has been based on the Berry curvature, which produces a fictitious magnetic field 
in the presence of dipole-dipole or Dzyalloshinzkii-Moriya (DM) interactions, both associated with spin-orbit (SO) coupling \cite{Okamoto17, Liu20}.  
Because it was borrowed from the theory of electronic structure \cite{Chang96,Sundaram99,Xiao10}, the Berry phase
is usually formulated in a semiclassical language.
For a Bloch function $\vert u_n(\vk )\rangle $ with energy $\epsilon_n(\vk )=\hbar \omega_n(\vk )$,
the Berry curvature 
\begin{equation}
{\bf \Omega}_n (\vk)=\frac{i}{2 \pi} 
\biggl\{ \frac{\partial }{\partial \vk } \times
\langle u_n(\vk ) \vert \frac{\partial }{\partial \vk }\vert u_n(\vk) \rangle \biggr\}
\label{EqBerry}
\end{equation}
of a ferromagnetic (FM) insulator
requires that a magnon wavepacket centered at $\vr_c $ obeys the equation of motion \cite{Chang96,Sundaram99,Xiao10}
\begin{equation}
\frac{d\vr_c}{dt }= \frac{\partial \epsilon_n(\vk )}{\hbar \,\partial \vk }-\frac{d \vk }{ dt} \times {\bf \Omega}_n(\vk ).
\end{equation}
Therefore, the Berry curvature causes the wavepacket to bend away from the expected direction $\partial \epsilon_n(\vk )/\partial \vk$ for a free magnon 
with ${\bf \Omega}_n(\vk )=0$.   

Prior to the first observation of the magnon Hall effect in the FM insulator Lu$_2$V$_2$O$_7$ \cite{Onose2010}, 
it was predicted by Katsura {\it et al.} \cite{Katsura2010} based on a Kubo formula for the 
temperature dependence of the thermal conductivity $\kappa^{xy}(T)$.
This Kubo formula involves an integral of the Berry curvature $\Omega_{nz}(\vk )$ perpendicular to the sample over the first Brillouin zone (BZ):
\begin{equation}
\kappa^{xy}(T)=-\frac{k_{\rm B}^2 T}{2\pi\hbar }\sum_n \int_{BZ} d^2k\, c_2\bigl(\rho(\epsilon_n(\vk ))\bigr)\,\Omega_{nz}(\vk ),
\label{kxy}
\end{equation}
where $\rho (\epsilon )=1/(\exp (\epsilon /k_{\rm B}T)-1)$ is the Boltzmann distribution with zero chemical potential
for magnons and $c_2(\rho )$ is defined in Section V.  The above expression includes the 
effects of the magnon edge currents travelling around the sample
as well as the effects of the magnon wavepacket ``self-rotation" due to its orbital motion \cite{Mat11a,Mat11b}.
The prediction and subsequent observation of the magnon Hall effect based on the Berry curvature
was one of the great early achievements in the field of magnonics.  

Because of SO coupling, magnons carry both spin and orbital angular momentum.
By constructing a Lagrangian that produces the correct equation of motion for the magnetization
${\bf M}_i=2\mb \vS_i$ at site $i$,
Tsukernik {\it et al.} \cite{Tsukernik66, Garmatyuk68} demonstrated that the orbital angular momentum (OAM) of magnons
along $\vz $ can be written in terms of
Bloch functions as 
\begin{equation}
O_n (\vk) =-\frac{i \hbar}{2} \biggl\{\vk \times \langle u_n(\vk )\vert \frac{\partial }{\partial \vk }\vert  u_n(\vk ) \rangle \biggr\}\cdot \vz. 
\label{EqOAM}
\end{equation}
But Tsukernik and coworkers failed to realize that the OAM defined above is not directly observable because,
unlike the Berry curvature ${\bf \Omega}_n(\vk)$, $O_n(\vk )$ is not gauge invariant \cite{Fukui2005}.  

This can be seen by expanding the spin Hamiltonian $H$ to 
second order in powers of the 
deviation of the spin operators $\vS_i$ from their equilibrium values. 
Then the Bloch functions $\vert u_n(\vk )\rangle $ 
satisfy the eigenvalue equation
\begin{equation}
H_2\vert u_n(\vk )\rangle = \epsilon_n(\vk )\vert u_n (\vk )\rangle .
\label{scev}
\end{equation}
Because $H_2$ is translationally invariant, a new Bloch function obtained by the transformation
\begin{equation}
\vert u_n(\vk )\rangle \rightarrow \vert u_n(\vk )\rangle \, e^{-i\lambda_n (\vk )}
\label{btr}
\end{equation}
also satisfies the above eigenvalue equation.
Under this gauge transformation, the Berry curvature ${\bf \Omega}_n (\vk)$ of Eq.\,(\ref{EqBerry}) remains unchanged but the OAM 
of Eq.\,(\ref{EqOAM}) changes to 
\begin{eqnarray}
O_n(\vk )&&\rightarrow O_n(\vk )+\frac{\hbar }{2}\biggl( k_x \frac{\partial }{\partial k_y}-k_y \frac{\partial }{\partial k_x}\biggr)\lambda_n(\vk ) \nonumber \\
&&=O_n(\vk )+\frac{\hbar }{2} \frac{\partial }{\partial \phi  }\lambda_n(k,\phi ),
\label{Lztr}
\end{eqnarray}
where the gauge $\lambda_n(\vk )=\lambda_n(k,\phi )$ depends only on the band index $n$ and the two-dimensional wavevector $\vk =(k,\phi )$.
Quantities like $O_n(\vk )$ that depend on a gauge $\lambda_n(\vk )$ cannot be physically observed \cite{Fukui2005}.
However, the average of $O_n(\vk )$ over all angles $\phi $ 
\begin{equation}
F_n(k)= \int_0^{2\pi }\frac{d\phi }{2\pi } \, O_n(\vk )
\label{lama}
\end{equation}
does not depend on the gauge $\lambda_n(\vk )$ \cite{Fishman23b}.  
Therefore, $F_n(k)$ can be physically observed.

\begin{figure}
\begin{center}
\includegraphics[width=8cm]{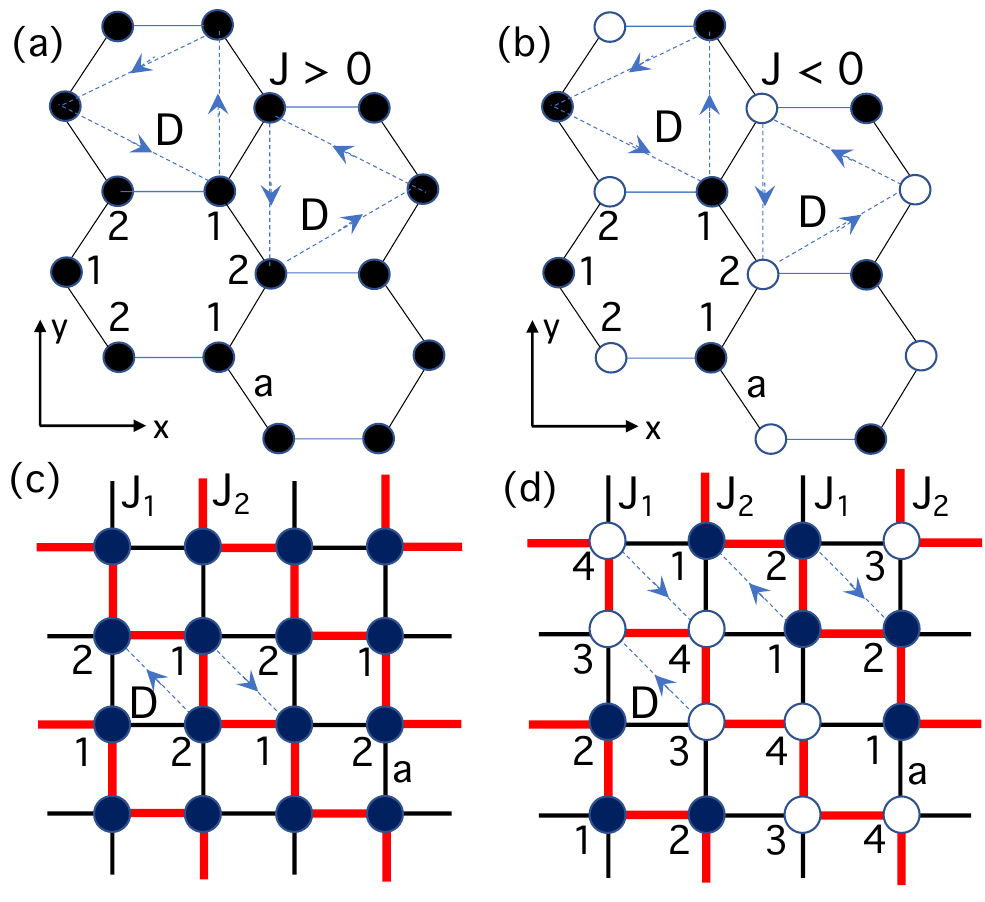}
\end{center}
\caption{Case studies:  HC lattices with (a) FM interaction $J>0$ and (b) AF interaction $J<0$ with two sites in the 
magnetic unit cell each;  ZZ lattices with (c) FM interactions $0 < J_1 < J_2$ and two sites in the magnetic unit cell, and
(d) AF interaction $J_1<0$ and FM interaction $J_2>0$ and four sites in the magnetic unit cell.  
The DM interaction $D$ and its orientation is shown by the dashed line.  Up spins are solid circles and down spins are 
empty circles.}
\label{Fig1}
\end{figure}

Regarding the magnon Hall effect as an indirect observation of the magnon OAM, then direct observation of the magnon OAM through the 
angular average $F_n(k)$ should be possible by coupling magnons to other particles and 
quasiparticles that carry OAM.  For example, magnons may couple to chiral phonons \cite{Zhu18} in crystals with broken inversion symmetry.  
High-energy electron beams separated into orbital components by a grating \cite{McMorran17} may also couple directly to the magnon OAM.

In this paper, we demonstrate that $F_n(k)$ is the {\it only} observable component of the OAM. 
We then evaluate $F_n(k)$ for FM honeycomb (HC) and square zig-zag (ZZ) lattices with DM interaction $D$.  
Along with their antiferromagnetic (AF) counterparts, these lattices are sketched in Fig.\,1.  Each model is described by the 
general Hamiltonian 
\begin{eqnarray}
H&=&-\frac{1}{2} \sum_{i,j} J_{ij}\,\vS_i \cdot \vS_j -D\sum_{i,j}(\vS_i\times \vS_j)\cdot \vz \nonumber \\
&-& K\sum_i (\vS_i \cdot \vz )^2,
\end{eqnarray}
with the DM interaction $-D(\vS_i\times \vS_j)\cdot \vz $ oriented along bond $i,j$ with spin $\vS_j$ at the endpoint 
and spin $\vS_i$ at the starting point of the arrow in Fig.\,1.  
For all four models, easy-axis anisotropy $K$ along $\vz $ is required to keep the DM interaction from tilting the spins away 
from the $z$ axis.  

In all four lattices, the DM interactions only act between sites $r$ of the same kind in each 
magnetic unit cell.  For the FM lattices of Figs.\,1(a) and (c), the DM interactions act between spins of type 1 or of type 2.
Similarly for the AF HC lattice in Fig.\,1(b).  For the AF ZZ lattice in Fig.\,1(d), the DM interactions act between spins of type 
1, 2, 3, or 4.  For both HC and ZZ lattices, DM interactions are allowed by broken inversion 
symmetry.  For the HC lattices, inversion symmetry is broken by lattice topology.  For the ZZ lattices, inversion 
symmetry is broken by the different environment to either side of an exchange path with $\vert J_2/J_1\vert \ne 1$.
By contrast, the environment around the midway point between neighboring spins is inversion symmetric on both lattices. 

OAM was earlier predicted \cite{Fishman22} to appear in the two ZZ lattices of Figs.\,1(c) and (d) with $\vert J_2/J_1\vert \ne 1$ but $D=0$.  
Unfortunately, OAM without DM interactions
cannot be observed due to its lack of gauge invariance \cite{Fishman23b}.  In this paper, we find that a FM ZZ lattice with a nonzero 
DM interaction $D\ne 0$ (see Fig.\,1(c)) creates a new class of materials where the effects of OAM are observable.
Due to the inequivalent DM interactions on either side of each bond,
the FM ZZ lattice then circumvents the ``no-go" theorem of Refs.\,\cite{Katsura2010} and \cite{Ideue12}, 
which is based on the edge sharing of equivalent cells.
However, because the magnon bands are always degenerate 
along the upper left and lower right boundaries of the BZ with $k_x -k_y = \pm \pi/a $, 
the Chern numbers $C_n$ of the magnon bands are not well defined.

The degeneracy of the FM ZZ bands along the upper left and lower right boundaries of the BZ can be 
lifted by allowing the FM exchange interaction $J_n$ along the 
$x$ and $y$ axis ($J_{nx}$ and $J_{ny}$) to be slightly different.
The Chern numbers of the magnon bands for the anisotropic FM ZZ model are then well defined and given by 0 or $\pm 1$.

There are several purposes served by this paper.  The next section extends earlier results for the angular-averaged OAM
$F_n(k)$.  By studying both the FM HC and ZZ lattices, we
show that formal results for $F_n(k)$ and the Berry curvature ${\bf \Omega}_n(\vk )$ for the two models are quite similar.
This paper also provides new results for the tiling of the OAM over all $\vk $ for both models.  Results for 
the FM ZZ model are completely new and presented here for the first time.

This paper is divided into six sections.  Section II demonstrates that $F_n(k)$ is the {\it only} component of 
$O_n(k)$ that is physically observable.  Sections III reviews results for the FM HC model.  Section IV 
discusses the FM ZZ model while Section V treats the anisotropic model with $J_{1y}/J_{1x}\ne 1$.
Section VI compares the predicted magnon Hall effects of the 
FM HC and ZZ models and contains a conclusion.  To focus attention on the 
FM cases where OAM can be observed, we treat the AF HC and ZZ lattices in Appendices B and C.  
While the FM lattices exhibit nonzero values of $F_n(k)$ when $D\ne 0$, the AF lattices do not:  $F_n(k)=0$ even when 
$D\ne 0$.  Appendix D presents results for the edge modes produced by the anisotropic FM ZZ model.
Since dipole-dipole interactions \cite{Okamoto17} are neglected in this paper, 
DM interactions provide the only source of SO coupling.

\section{Components of the OAM}

To set the stage for the results provided in the next two sections, we briefly review the spin-wave (SW) formalism for the OAM and Berry curvature, specializing
to collinear spin systems.  Rotated into the local spin reference frame with ${\bar {\vz}}_i$ pointing along the spin direction, the spins 
$\bar {\bf S}_i$ are given in terms of the Boson SW creation and 
annihilation operators $a_i$ and $a_i^{\dagger }$ by ${\bar S}_{iz}=S-a_i^{\dagger }a_i$,
${\bar S}_{i+}=S_{ix }n_{iz}+iS_{iy}=\sqrt{2S}a_i$, and ${\bar S}_{i-}=S_{ix}n_{iz}-iS_{iy}=\sqrt{2S}a_i^{\dagger}$, 
where $n_{iz}=1$ for up spins and $n_{iz}=-1$ for down spins.
Then the Hamiltonian $H$ can be expanded in powers of the Fourier transformed SW operators $a_{\vk }^{(r)}$ and $a_{\vk }^{(r)\dagger }$ 
($r$ is one of the $M$ sites in the magnetic unit cell) as $H= E_0 + H_2 + \ldots $ with second-order Hamiltonian
\begin{equation}
H_2={\sum_\vk }' {\bf v}_{\vk}^{\dagger }\cdot \underline{{\it L}}(\vk )\cdot {\bf v}_{\vk },
\label{defL}
\end{equation}
where the prime indicates that the summation over $\vk $ is restricted to the 
first BZ of the magnetic unit cell.
The $2M$-dimensional vector operators
\begin{equation}
{\bf v}_{\vk } =(a_{\vk }^{(1)},a_{\vk }^{(2)},\ldots  a_{\vk }^{(M)},a_{-\vk}^{(1)\dagger },a_{-\vk }^{(2)\dagger },\ldots a_{-\vk }^{(M)\dagger })
\end{equation}
satisfy 
$[{\bf v}_{\vk },{\bf v}^{\dagger }_{\vk'}] =\underline{N}\,\delta_{\vk ,\vk'}$,
where $\underline{N}$ is defined in terms of the $M$-dimensional identity matrix $\underline{I}$ by
\begin{equation}
\underline{N} =
\left(
\begin{array}{cc}
\underline{I} &0 \\
0 & -\underline{I} \\
\end{array} \right).
\label{defn}
\end{equation}
The $2M\times 2M$ matrix $\uL (\vk)$ can be compactly written
\begin{equation}
\uL (\vk ) =\left( \begin{array}{cc}\uP (\vk ) & \uQ (\vk ) \\
\uQp (\vk )& \uPp (\vk)  
\end{array} \right) ,
\end{equation}
where $\uP (\vk )$, $\uQ (\vk )$, $\uPp (\vk)$, and $\uQp (\vk )$ are $M \times M$ matrices.  Because
$\uL (\vk )$ is Hermitian, 
\begin{equation}
\label{up}
\uPp (\vk ) = \uP (-\vk )^{\star },
\end{equation}
\begin{equation}
\label{uq}
\uQp (\vk ) = \uQ (-\vk )^{\star }.
\end{equation}
These relations will prove useful in the following two sections and in Appendices B through D.

Within the quantum SW notation, the semiclassical eigenvalue relation of Eq.\,(\ref{scev})
is replaced by 
\begin{equation}
\underline{\Lambda} (\vk )\cdot\underline{X}^{-1}(\vk ) = \hbar \omega_n(\vk )\,\underline{X}^{-1 }(\vk ),
\label{egv}
\end{equation}
where ${\underline \Lambda}(\vk ) = {\underline N}\cdot \underline{L}(\vk )$ is non-Hermitian.
Hence, the 
Bloch functions $\vert u_n (\vk )\rangle $ are replaced by the complex matrices $X^{-1 }(\vk )_{rn}$,
which can be considered the $n$th eigenfunctions of the $2M \times 2M$ energy matrix ${\underline{\Lambda}}(\vk )$.
In the quantum SW language, the Berry curvature and OAM are given by 
\begin{eqnarray}
&&{\bf \Omega}_n (\vk )= \frac{i}{2\pi}\sum_{r=1}^M\biggl\{ 
\frac{\partial X^{-1}(\vk )_{rn}^*}{\partial \vk } \times \frac{\partial X^{-1}(\vk )_{rn}}{\partial \vk } \nonumber \\
&&- \frac{ \partial X^{-1}(\vk)_{r+M,n}^*}{\partial \vk} \times  \frac {\partial X^{-1}(\vk )_{r+M,n}}{\partial \vk } \biggr\},
\label{Bfdef}
\end{eqnarray}
\begin{eqnarray}
&&O_n(\vk )= \frac{\hbar}{2}\sum_{r=1}^M\Bigl\{ 
X^{-1}(\vk )_{rn}\,\lk \, X^{-1}(\vk )_{rn}^* \nonumber \\
&&- X^{-1}(\vk)_{r+M,n}\,\lk \, X^{-1}(\vk )^*_{r+M,n}\Bigr\},
\label{Lzdef}
\end{eqnarray}
where
\begin{equation}
\lk = -i \biggl( k_x\frac{\partial }{\partial k_y}-k_y\frac{\partial}{\partial k_x}\biggr)
\label{lk}
\end{equation}
is the OAM operator.
The normalization condition for the Bloch functions $\langle u_n(\vk )\vert u_n(\vk )\rangle =1$ is then replaced by
the condition
\begin{equation}
\sum_{r=1}^M \Bigl\{ \vert X^{-1}(\vk )_{rn}\vert^2 - \vert X^{-1}(\vk )_{r+M,n}\vert^2\Bigr\}=1
\label{sumx}
\end{equation}
for the complex matrices $X^{-1}(\vk )_{rn}$ and $X^{-1}(\vk )_{r+M,n}$.

With the Berry curvature defined above, the Chern number for band $n$ is given by
\begin{equation}
C_n=\int_{BZ} d^2k\, \Omega_{nz}(\vk ),
\end{equation}
where $\vk $ is integrated over the first BZ \cite{Shindou13}.  A customary factor of $1/2\pi $ is missing from this expression for $C_n$ because it is
included in Eqs.\,(\ref{EqBerry}) and (\ref{Bfdef}) for the Berry phases.  The Chern number $C_n$ takes an integer value so long as the magnons in band $n$ 
are nondegenerate, i.e. disconnected from all other magnons in frequency for all $\vk $.  A nonzero Chern number is physically
associated with edge modes \cite{Mat11a, Mat11b,Mong11,Zhang13,Mook14a} whose dispersion bridges the gap between bulk magnon bands.
Since the sum of Berry curvatures over all bands vanishes, $\sum_n C_n=0$.
 
In the quantum language, 
each eigenfunction $X^{-1 }(\vk )_{rn}$ can be multiplied by an arbitrary phase factor 
so that
\begin{equation}
X^{-1}(\vk)_{rn }\rightarrow X^{-1}(\vk )_{rn}\,e^{-i\lambda_n(\vk )},
\label{gtr}
\end{equation}
where the gauge $\lambda_n(\vk )$ may depend on band index $n$ and $\vk =(k,\phi )$ but not on site $r$.  
Of course, $\lambda_n(k,\phi )$ must also be a single-valued function of $\phi $ so that $\lambda_n(k,0)=\lambda_n(k,2\pi)$.
Under a gauge transformation, the OAM changes by
\begin{equation}
O_n(\vk )\rightarrow O_n(\vk )+\frac{\hbar }{2} \frac{\partial }{\partial \phi  }\lambda_n(k,\phi ),
\label{rhso}
\end{equation}
in agreement with the semiclassical expression of Eq.\,(\ref{Lztr}).

Now expand $O_n(k,\phi )$ in powers of $\cos l\phi $ and $\sin l\phi $
so that
\begin{equation}
O_n(k,\phi )=\sum_{l=0} \Bigl\{ A_{ln}(k) \cos l\phi + B_{ln}(k) \sin l\phi \Bigr\}.
\end{equation}
Following a gauge transformation, the $rhs$ of Eq.\,(\ref{rhso}) becomes
\begin{eqnarray}
&&\sum_{l=0} \Bigl\{ A_{ln}(k) \cos l\phi + B_{ln}(k) \sin l\phi \Bigr\} +\frac{\hbar }{2} \frac{\partial }{\partial \phi  }\lambda_n(k,\phi )\nonumber \\
&&= A_{0n}(k) + \sum_{l=1} \Bigl\{ A_{ln}(k) \cos l\phi + B_{ln}(k) \sin l\phi \Bigr\} \nonumber \\
&&+\frac{\hbar }{2} \frac{\partial }{\partial \phi  }\lambda_n(k,\phi )=A_{0n}(k),
\end{eqnarray}
where we have set 
\begin{eqnarray}
\lambda_n(k,\phi )&&= -\frac{2}{\hbar } \sum_{l=1} \frac{1}{l} \Bigl\{ A_{ln}(k) \sin l\phi \nonumber \\
&&-B_{ln} (k) \cos l\phi \Bigr\}.
\end{eqnarray}
The $l=0$ component cannot be included on the $rhs$ because it would produce a term 
proportional to $\phi $, violating the assumption that $\lambda_n(k,0)=\lambda_n(k,2\pi)$.
Hence, the appropriate gauge $\lambda_n(k)$ can be used to remove all components of the OAM except for 
\begin{equation}
A_{0n}(k) = F_n(k) \equiv \int_0^{2\pi }\frac{d\phi }{2\pi } \, O_n(\vk ).
\end{equation}
Not only does this prove that $F_n(k)$ is observable but it also demonstrates that $F_n(k)$ is the {\it only} observable component of the OAM.

In the absence of DM interactions, the OAM $O_n(\vk )$ is an odd function of $\vk $ so that
$O_n(\vk )=-O_n(-\vk )$ and $F_n(k )=0$.  When the DM interaction $D$ enters $O_n(\vk )$ linearly, then $O_n(\vk )=O_n(-\vk )$ is an even function 
of $\vk $ and $F_n(k)$ can be nonzero.  More generally, we expand $O_n(\vk )$ in powers of the DM interaction as
\begin{eqnarray}
O_n(\vk ) &=& O_n^{(0)}(\vk ) + D\,O_n^{(1)}(\vk )+D^2\, O_n^{(2)}(\vk )\nonumber \\
&&+D^3\, O_n^{(3)}(\vk ) + \ldots.
\end{eqnarray}
Then, the even components $O_n^{(2m)}(\vk )=-O_n^{(2m)}(-\vk )$ are odd in $\vk $ and the odd components $O_n^{(2m+1)}(\vk )=O_n^{(2m+1)}(-\vk )$ are even in $\vk $.  
Of course, only the odd components
$O_n^{(2m+1)}(\vk )$ contribute to $F_n(k)$.
If we then write
\begin{equation}
O_n(\vk )=O_n^{({\rm odd})}(\vk )+O_n^{({\rm even})}(\vk ),
\end{equation}
only $O_n^{({\rm even})}(\vk )$ (containing terms of order $D^{2m+1}$) contributes to 
the physically measureable $F_n(k)$.

Since only wavevectors $\vk $ within the first BZ of the magnetic unit cell enter Eq.\,(\ref{defL}), we use periodic boundary conditions 
to evaluate the integral over angles $\phi $ in $F_n(k)$ when required to translate wavevectors $\vk $ outside the first BZ to wavevectors inside the first BZ.
Alternatively, $F_n(k)$ can be evaluated by tiling all of $\vk $ space with the first BZ of $O_n^{({\rm even})}(\vk )$.  For the FM ZZ lattice,
Appendix A shows that the resulting pattern for 
$O_n^{({\rm tiled})}(\vk )$ is both periodic in $\vk $ and continuous as a function of $\vk $ at the zone boundaries.  
We shall give examples of the tiling procedure for the FM HC and
ZZ models in the following sections.  Nevertheless, bear in mind that $O_n^{({\rm tiled})}(\vk )$ is not unique and is just
a tool to evaluate the physically observable quantity $F_n(k)$.

\section{FM Honeycomb lattice}

Most details of the solution for the FM HC lattice sketched in Fig.\,1(a) with exchange $J>0$ and DM interaction $D$
between like sites were previously provided in Ref.\,\cite{Fishman23b}.  
The $4\times 4$ matrix $\underline{L}(\vk )$ defined by Eq.\,(\ref{defL}) 
is given by
\begin{equation}
\underline{L}(\vk ) =
\frac{3JS}{2} \left(
\begin{array}{cccc}
A^-_{\vk }  & -\Gamma_{\vk}^* & 0 & 0 \\
-\Gamma_{\vk } & A^+_{\vk }  & 0 & 0 \\
0 & 0 & A^+_{\vk } & - \Gamma_{\vk }^*\\
0 & 0 &-\Gamma_{\vk } & A^-_{\vk }\\
\end{array} \right),
\end{equation}
where $A^{\pm }_{\vk }= 1 \pm d\, \Theta_{\vk } + \kappa $, 
$d=-2D/3J$, $\kappa = 2K/3J$,
\begin{equation}
\Theta_{\vk} = 4\cos(3k_xa/2) \sin(\sqrt{3} k_ya/2)-2\sin(\sqrt{3}k_ya),
\end{equation}
and
\begin{equation}
\Gamma_{\vk} =\frac{1}{3}\Bigl\{ e^{ik_xa}+2e^{-ik_xa/2} \cos(\sqrt{3}k_ya/2) \Bigr\}.
\end{equation}
We caution the reader that matrix element $A^{\pm }_{\vk }$, DM parameter $d$, and anisotropy parameter $\kappa $ shall be 
defined differently for the FM ZZ lattice in the next section.
Since $\Theta_{-\vk }=-\Theta_{\vk }$
and $\Gamma_{-\vk }=\Gamma_{\vk }^*$, it can be easily shown that the upper and lower quadrants of $\uL (\vk )$ satisfy Eq.\,(\ref{up})
or that $\uPp (\vk ) = \uP (-\vk )^{\star }$.

\begin{figure}
\begin{center}
\includegraphics[width=9cm]{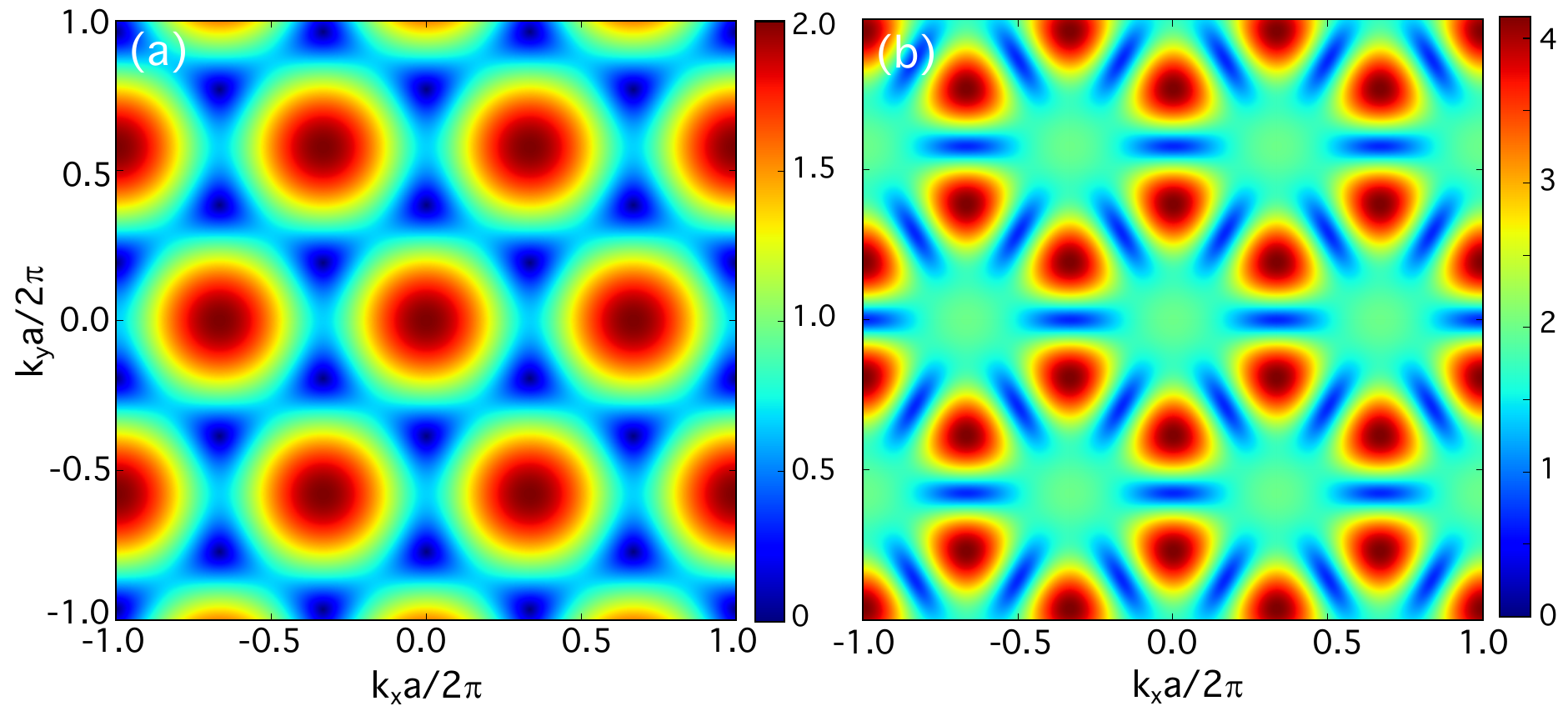}
\end{center}
\caption{The normalized gap $\delta (\vk )=2\eta_{\vk }$ for the FM HC lattice versus $\vk $ for $d$ = (a) 0 and (b) $-0.4$.}
\label{Fig2}
\end{figure}

Magnon energies for bands 1 and 2 are given by
\begin{equation}
\hbar \omega_1(\vk)=3JS ( 1  - \eta_{\vk } +\kappa )
\label{om1},
\end{equation}
 \begin{equation}
\hbar \omega_2(\vk)=3JS ( 1  + \eta_{\vk } +\kappa ),
\label{om2}
\end{equation}
where $\eta_{\vk }=\sqrt{\vert \Gamma_{\vk }\vert^2 +(d\, \Theta_{\vk })^2}$.
Notice that these energies
are simply shifted by the easy-axis anisotropy $\kappa $.
The magnon band gap is given by
\begin{equation}
\hbar \Delta \omega (\vk ) = \hbar (\omega_2(\vk )-\omega_1(\vk )) =  6JS\eta_{\vk }.
\end{equation}
The normalized gap
\begin{equation}
\delta (\vk ) \equiv \frac{\hbar \Delta \omega (\vk )}{6JS}=2\eta_{\vk }
=2\sqrt{\vert \Gamma_{\vk }\vert^2 +(d\, \Theta_{\vk } )^2}
\end{equation}
is plotted versus $\vk $ for $d=0$ and $-0.4$ in Fig.\,2.  When $d=0$, the magnon gap vanishes in triangular $\vk $-space regions at the corners of the BZ.  When $d=-0.4$,
the smallest normalized gap $\delta (\vk )$ is 2/3.  In fact, any nonzero $d$ introduces a gap such that the $\delta (\vk )> 0 $ and modes 1 and 2 are distinct for all $\vk $.

\begin{figure}
\begin{center}
\includegraphics[width=8cm]{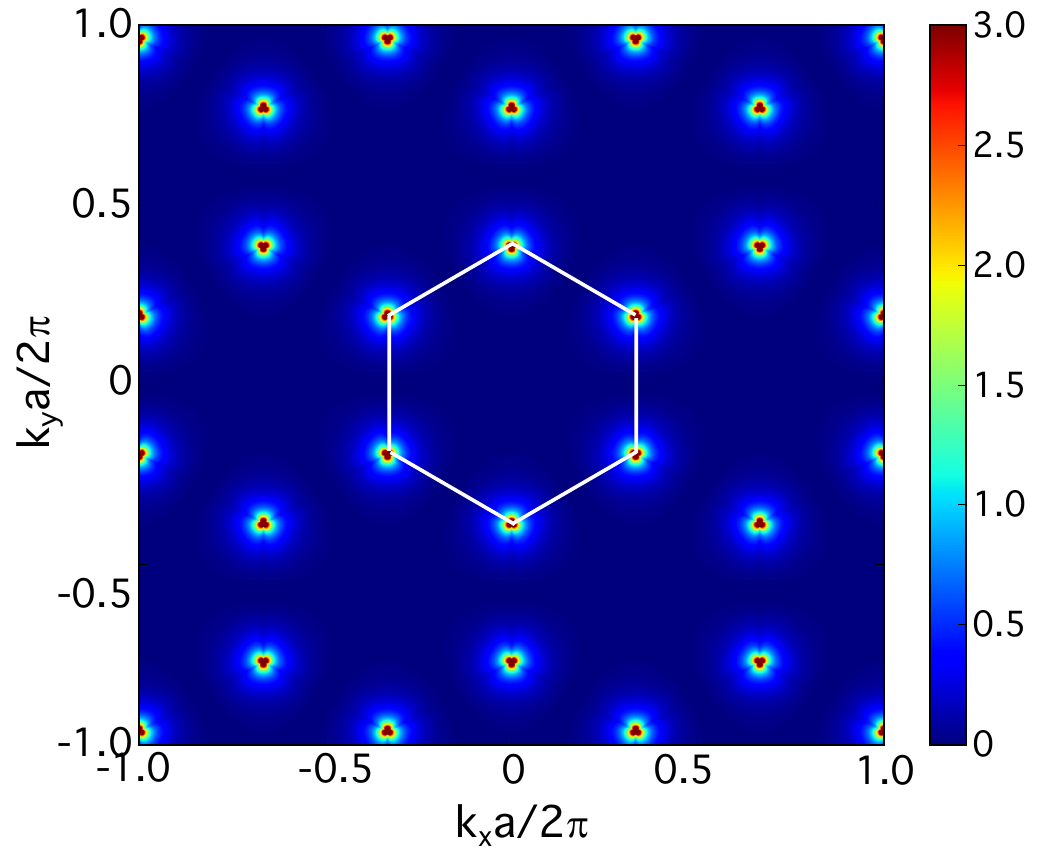}
\end{center}
\caption{The pattern $O_1^{({\rm tiled})}(\vk )/\hbar $ for the FM HC lattice with $d=-0.1$.  The first BZ of the magnetic unit cell is denoted by the solid white lines.}
\label{Fig3}
\end{figure}

\begin{figure}
\begin{center}
\includegraphics[width=7cm]{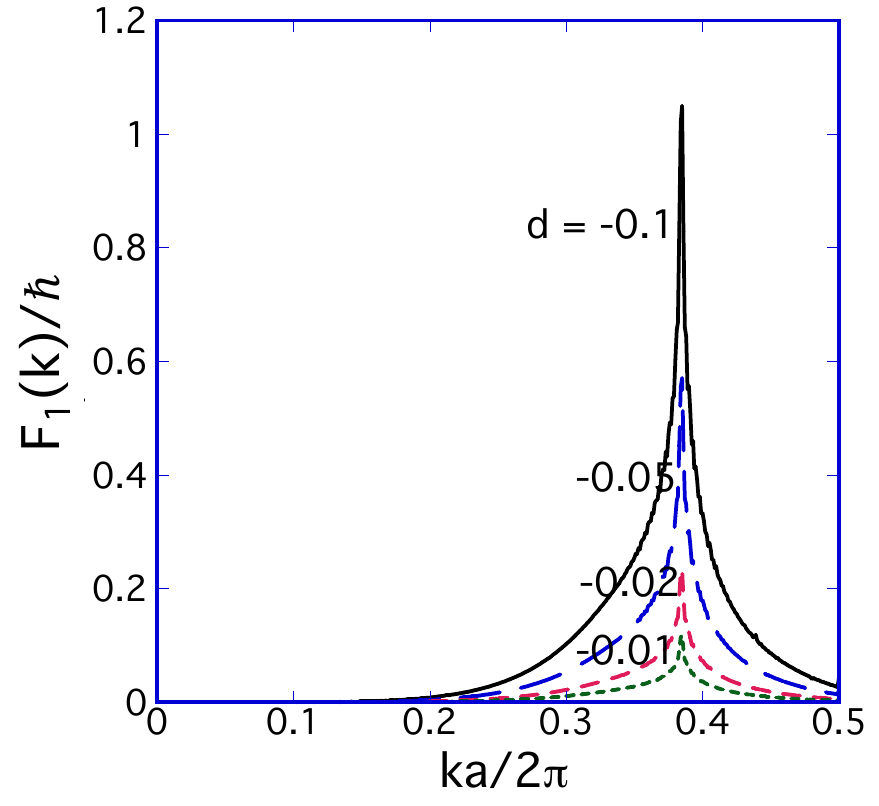}
\end{center}
\caption{The function $F_1(k)/\hbar $ evaluated using $O_1^{({\rm tiled})}(\vk )/\hbar $ for the FM HC lattice with $d=-0.01$, $-0.02$, $-0.05$, and $-0.1$.}
\label{Fig4}
\end{figure}

Using a particularly simple 
form for the gauge, we earlier found \cite{Fishman23b} (correcting a minus sign) that
\begin {equation}
O_1(\vk )=\frac{\hbar }{4} \biggl\{ 1 - \frac{d\,\Theta_{\vk } }{\eta_{\vk }} \biggr\}  \frac{\Gamma_{\vk}}{ \vert \Gamma_{\vk }\vert }\,
 \lk \,  \frac{\Gamma_{\vk }^*}{\vert \Gamma_{\vk }\vert },
 \label{lzl}
\end{equation}
\begin {equation}
O_2 (\vk)=\frac{\hbar }{4} \biggl\{ 1 +\frac{d\,\Theta_{\vk } }{\eta_{\vk }} \biggr\}  \frac{\Gamma_{\vk}}{ \vert \Gamma_{\vk }\vert }\,
 \lk \,  \frac{\Gamma_{\vk }^*}{\vert \Gamma_{\vk }\vert }.
 \label{lzu}
 \end{equation}
Unlike the mode frequencies, however, solutions for the OAM are not unique.  Since
the $d=0$ portion of the OAM is not observable, the first terms in the brackets
of Eqs.\,(\ref{lzl}) and (\ref{lzu}) can be neglected.  Because $\eta_{\vk }$ is an even function of $d$, we then find 
\begin {equation}
O_1^{({\rm even})}(\vk )=-O_2^{({\rm even})}(\vk )=-\frac{d\,\hbar }{4}\frac{\Theta_{\vk } }{\eta_{\vk }} \frac{\Gamma_{\vk}}{ \vert \Gamma_{\vk }\vert }\,
 \lk \,  \frac{\Gamma_{\vk }^*}{\vert \Gamma_{\vk }\vert }
\label{lze}
\end{equation}
and
\begin{equation}
F_1(k) = -F_2(k)= -\frac{d\,\hbar }{4} \int \frac{d\phi }{2\pi } \frac{\Theta_{\vk }}{\eta_{\vk }}\frac{\Gamma_{\vk}}{ \vert \Gamma_{\vk }\vert }\,
 \lk \,  \frac{\Gamma_{\vk }^*}{\vert \Gamma_{\vk }\vert }.
\label{fze}
\end{equation}
Figure 3 uses $O_1^{({\rm even})}(\vk )/\hbar $ with $d=-0.1$ to construct $O_1^{({\rm tiled})}(\vk )/\hbar $.   Notice that the tiled pattern is both a periodic function of 
$\vk $ and a continuous function of $\vk $ at the boundaries of the first BZ, denoted by the solid lines.
The solutions for $F_1(k)/\hbar $ for $d$ running from $-0.01$ to $-0.1$ are plotted in Fig.\,4.  We find that $F_1(k)/\hbar $ grows quite rapidly with 
$d$ and peaks when $ka/2\pi = 2\sqrt{3}/9 = 0.385$, which coincides with the corners of the first BZ.  The value of $F_1(k)/\hbar $ at $ka/2\pi = 2\sqrt{3}/9$ diverges as
$\vert d\vert \rightarrow \infty $.

While Fig.\,3 suggests that the OAM is largest at the corners of the BZ and Fig.\,4 does not disallow that claim, we remind the reader that Fig.\,4 only states that the angular average
of the OAM is largest when $ka/2\pi$ intercepts the corners of the BZ.  It does {\it not} imply that the largest OAM lies for $\phi $ at the BZ corners.

\begin{figure}
\begin{center}
\includegraphics[width=9cm]{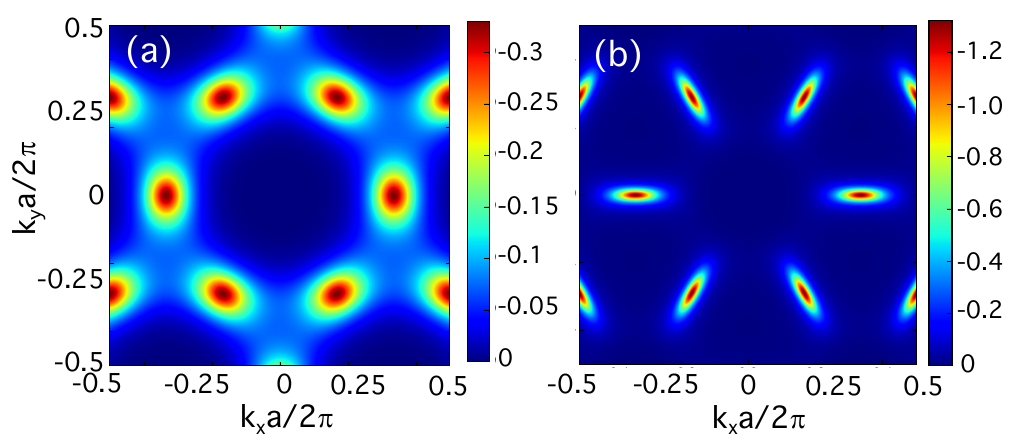}
\end{center}
\caption{The Berry curvature $\Omega_{1z}(\vk )$ of the FM HC lattice versus $\vk $ for $d=$ (a) $-0.1$ and (b) $-0.4$.}
\label{Fig5}
\end{figure}

Again correcting a minus sign, the Berry curvatures of the FM HC lattice are given analytically by 
\begin{eqnarray}
&&\Omega_{1z}(\vk ) =-\Omega_{2z}(\vk )\nonumber \\ 
&&=i\frac{d }{4\pi } \, \frac{\Gamma_{\vk}^*}{\vert \Gamma_{\vk }\vert } 
\Biggl\{ \frac{\partial \Theta_{\vk } /\eta_{\vk } }{\partial \vk } \times \frac{\partial \Gamma_{\vk }/\vert \Gamma_{\vk } \vert }{\partial \vk }  \Biggr\} \cdot \vz .
\label{bfhc}
\end{eqnarray}
For the lower band, $\Omega_{1z}(\vk )$ is plotted in Fig.\,5 for $d=-0.1$ and $-0.4$.  The Chern numbers $C_n$ of the lower and upper magnon 
bands are $-1$ and $+1$, respectively, for all $d<0$.  The Chern number is an integer due to the 
nonzero gap \cite{Shindou13} between the magnon modes for all $\vk $. 

\section{FM Zig-zag lattice}

The FM ZZ lattice is sketched in Fig.\,1(c) with exchange interactions $0 < J_1 < J_2$ and
DM interaction $D$ between like sites along $(1,-1)$.  Then
\begin{equation}
\underline{L}(\vk ) =
(J_1+J_2)S\left(
\begin{array}{cccc}
A^-_{\vk }  & -\Psi_{\vk}^* & 0 & 0 \\
-\Psi_{\vk } & A^+_{\vk }  & 0 & 0 \\
0 & 0 & A^+_{\vk } & -\Psi_{\vk }^*\\
0 & 0 &-\Psi_{\vk } & A^-_{\vk }\\
\end{array} \right),
\end{equation}
where $A^{\pm }_{\vk }= 1\pm d\, \tau_{\vk } +\kappa $, $d=-2D/(J_1+J_2)$, $\kappa = K/(J_1+J_2)$, $\tau_{\vk }=\sin (k_ya-k_xa)$,
and
\begin{equation}
\Psi_{\vk }=\frac{J_1\xik^*+J_2\xik }{2(J_1+J_2)}
\end{equation}
with $\xik =\exp(ik_xa) + \exp(ik_ya)$.   Using $\tau_{-\vk }=-\tau_{\vk }$
and $\Psi_{-\vk }=\Psi_{\vk }^*$, it is easy to verify that the upper and lower quadrants of $\uL (\vk )$ satisfy Eq.\,(\ref{up})
or that $\uPp (\vk ) = \uP (-\vk )^{\star }$.

The magnon energies are given by
\begin{equation}
\hbar \omega_1(\vk)=2(J_1+J_2)S ( 1  - \mu_{\vk } + \kappa )
\label{zzfm1},
\end{equation}
 \begin{equation}
\hbar \omega_2(\vk)=2(J_1+J_2)S ( 1  + \mu_{\vk } + \kappa ),
\label{zzfm2}
\end{equation}
where
$\mu_{\vk }=\sqrt{\vert \Psi_{\vk }\vert^2 +(d\, \tau_{\vk } )^2}$.  As for the FM HC lattice, the magnon bands are just shifted by $\kappa $.
The gap between the magnons is then given by
\begin{equation}
\hbar \Delta \omega (\vk ) = \hbar (\omega_2(\vk )-\omega_1(\vk )) =  4(J_1+J_2)S\mu_{\vk },
\end{equation}
with a normalized gap 
\begin{equation}
\delta (\vk ) \equiv \frac{\hbar \Delta \omega (\vk )}{2(J_1+J_2)S}=2\mu_{\vk }
=2\sqrt{\vert \Psi_{\vk }\vert^2 +(d\, \tau_{\vk } )^2}
\end{equation}
that only depends on $d$ and $r=J_2/J_1$.  The normalized gap
is plotted versus $\vk $ on the top two panels of Fig.\,6 for $d=0$ and $r= 8$ or 1.  
For $r > 1$, $\delta (\vk )=0$ at the upper left and lower right borders of the BZ, which is sketched by the rotated square.
For $r= 1$, $\delta (\vk )=0$ at all four borders of the BZ.  
The bottom two panels of Fig.\,6 plot $\delta (\vk )$ for the same values of $r$ with $d=-0.4$.  Even for $d\ne 0$, 
the gap $\delta (\vk )$ continues to vanish at the upper left and lower right boundaries of the BZ with $k_x-k_y = \pm \pi /a$
because $\tau_{\vk }=0$.  

\begin{figure}
\begin{center}
\includegraphics[width=9cm]{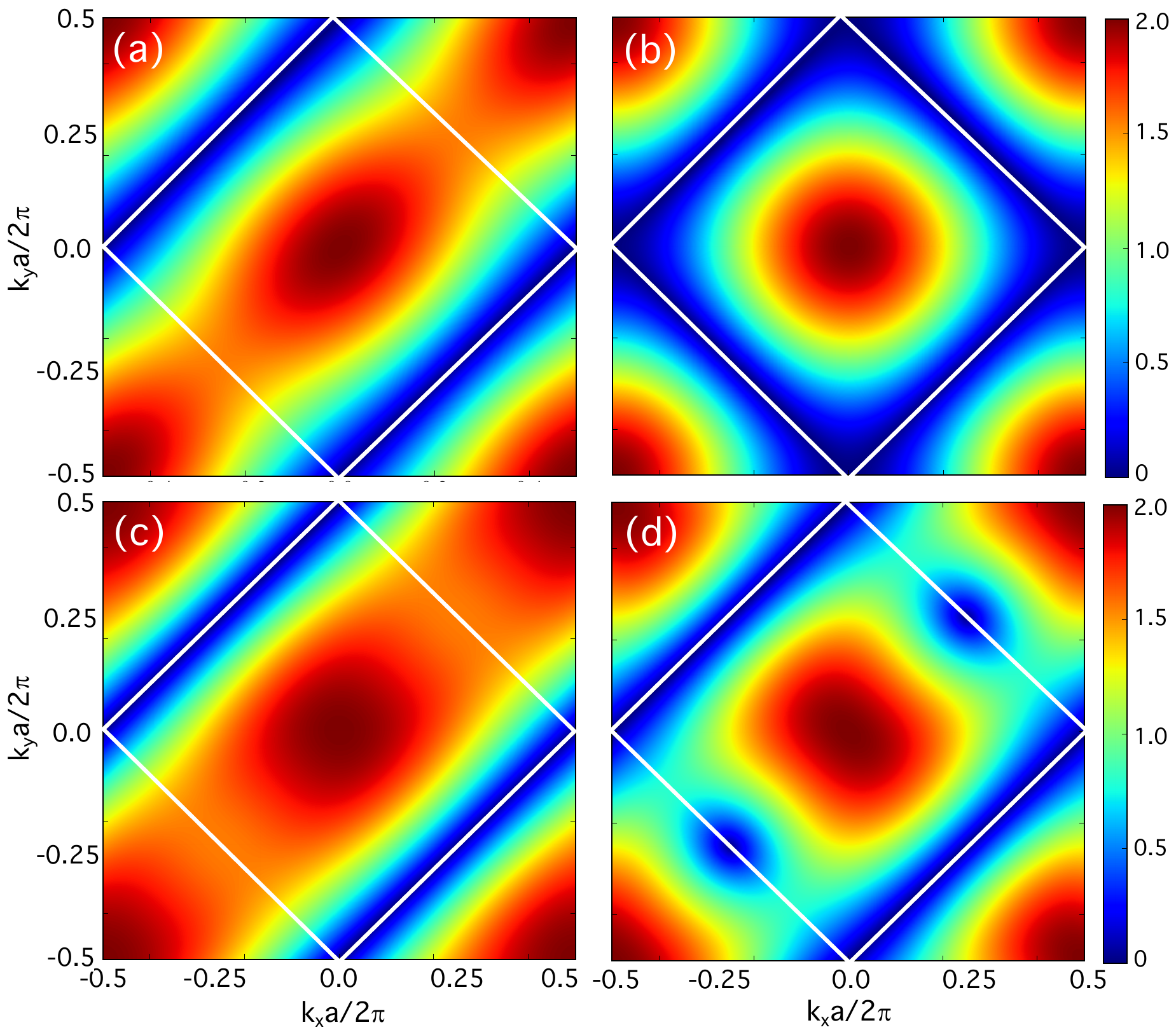}
\end{center}
\caption{The normalized gap $\delta (\vk )=2\mu_{\vk }$ between bands of the FM ZZ lattice evaluated using $d=0$ and $r$ = (a) 8 and (b) 1 on the top two
panels or $d=-0.4$ and $r$ = (c) 8 and (d) 1 on the bottom two panels.  The first BZ
of the magnetic unit cell is sketched by the solid lines.}
\label{Fig6}
\end{figure}

\begin{figure}
\begin{center}
\includegraphics[width=9cm]{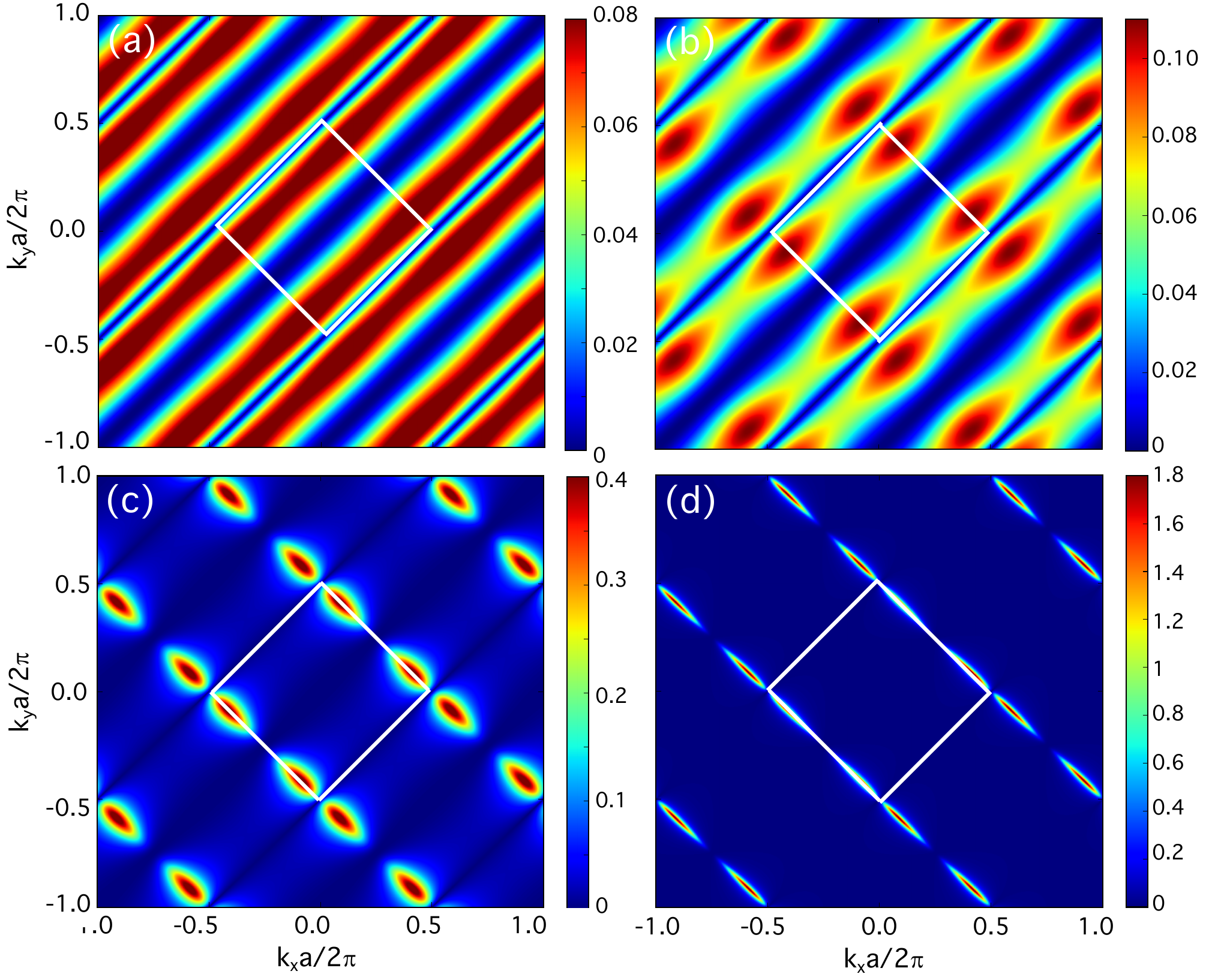}
\end{center}
\caption{The pattern $O_1^{({\rm tiled})}(\vk )/\hbar $ for the FM ZZ lattice with $d=-0.4$ and $r=$ (a) 50, (b) 8, (c) 1.5, and (d) 1.1.  
The first BZ of the magnetic unit cell is denoted by the solid white lines.}
\label{Fig7}
\end{figure}

The solutions for $O_n^{({\rm even})}(\vk )$ and $F_n(k)$ are formally quite similar to those for the FM HC lattice in Eq.\,(\ref{lze}) and (\ref{fze}):
\begin {equation}
O_1^{({\rm even})}(\vk )=-O_2^{({\rm even})}(\vk )=-\frac{d\,\hbar }{4}\frac{\tau_{\vk } }{\mu_{\vk }} \frac{\Psi_{\vk}}{ \vert \Psi_{\vk }\vert }\,
 \lk \,  \frac{\Psi_{\vk }^*}{\vert \Psi_{\vk }\vert }
\end{equation}
and
\begin{equation}
F_1(k) = -F_2(k)= -\frac{d\,\hbar }{4} \int \frac{d\phi }{2\pi } \frac{\tau_{\vk }}{\mu_{\vk }}\frac{\Psi_{\vk}}{ \vert \Psi_{\vk }\vert }\,
 \lk \,  \frac{\Psi_{\vk }^*}{\vert \Psi_{\vk }\vert }.
\end{equation}
The pattern $O_1^{({\rm tiled})}(\vk )/\hbar $ is plotted in Fig.\,7 for $d=-0.4$ and $r=50$, 8, 1.5, and 1.1.  
For $r=50$ or 8, $O_1^{({\rm tiled})}(\vk )/\hbar $
contains wide troughs of minima close to zero near avenues of maxima close to $0.08$ or $0.11\,\hbar $, both along $(1,1)$.  Narrow lanes of vanishing OAM appear
at the $k_x-k_y=\pm \pi /a$ boundaries of the BZ where the magnon bands are degenerate.  The OAM also vanishes in a lane along $(1,1)$ that crosses
$\vk =0$.  For $r=1.5$ and 1.1, peaked regions in $O_1^{({\rm tiled})}(\vk )/\hbar $ appear at the $k_x+k_y=\pm \pi /a$
boundaries of the BZ along $(1,-1)$.  These regions become increasingly narrow as $r\rightarrow 1$.

The observable function $F_n(k)/\hbar $ for the FM ZZ lattice is plotted in Fig.\,8 for $d=-0.4$ and these same four values of $r$. 
For $r=1.5$ and 1.1, the large maxima of $F_1(k)/\hbar $ at $ka/2\pi \approx 0.4$ are associated with the peaked regions of $O_1^{({\rm tiled})}(\vk )/\hbar $ 
at the lower left and upper right boundaries of the BZ ($k_x+k_y = \pm \pi /a$) in Fig.\,7.  
The observable portion of the OAM becomes increasingly narrow and disappears as $r \rightarrow 1$.

\begin{figure}
\begin{center}
\includegraphics[width=7cm]{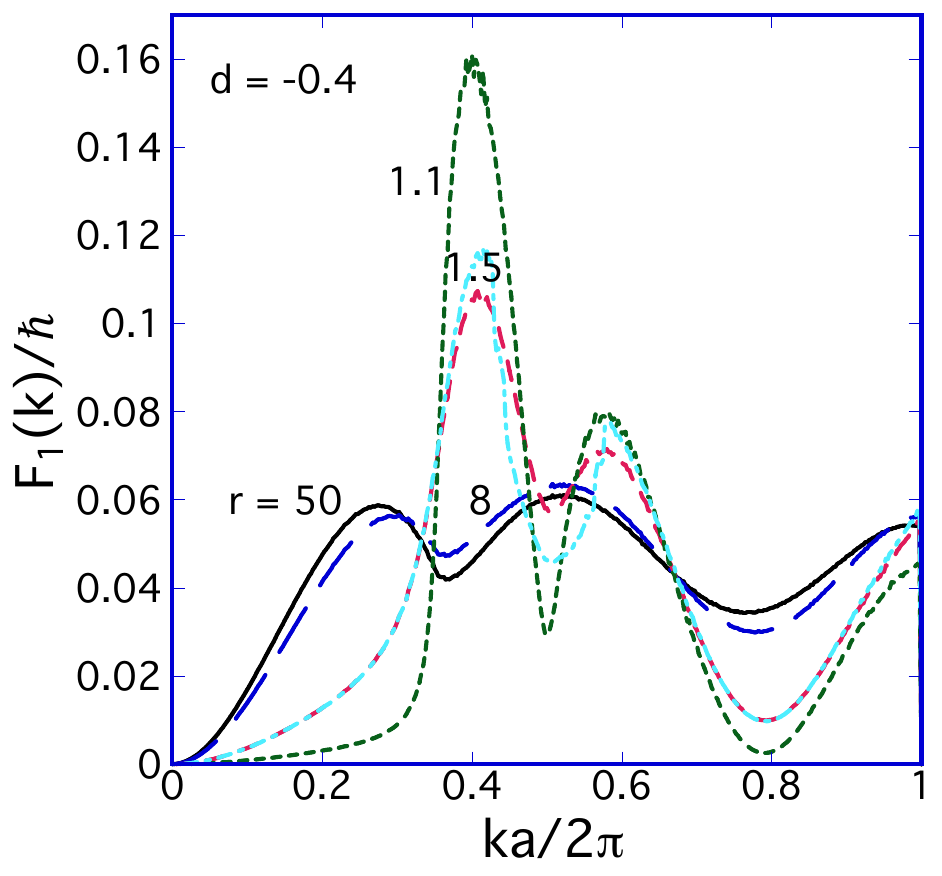}
\end{center}
\caption{The function $F_1(k)/\hbar $ evaluated using $O_1^{({\rm tiled})}(\vk )/\hbar $ for the FM ZZ lattice with $d=-0.4$ and 
$r$ = 50, 8, 1.5 (long-dash curve), and 1.1.  The short-dash curve for $r=1.5$ takes $J_{1y}/J_{1x}=1.1$ in the revised ZZ model.}
\label{Fig8}
\end{figure}

Since the magnon Hall effect may be observed in the FM ZZ lattice with DM interaction, we also provide results for its 
Berry curvature.  The Berry curvatures along $\vz$ may be written as 
\begin{eqnarray}
&&\Omega_{1z}(\vk ) =-\Omega_{2z}(\vk )\nonumber \\
&&= i\frac{d}{4\pi } \, \frac{\Psi_{\vk}^*}{\vert \Psi_{\vk }\vert } 
\Biggl\{ \frac{\partial \,\tau_{\vk } /\mu_{\vk } }{\partial \vk } \times \frac{\partial \,\Psi_{\vk }/\vert \Psi_{\vk } \vert }{\partial \vk }  \Biggr\} \cdot \vz ,
\label{bfzig}
\end{eqnarray}
which are formally similar to the expressions for the Berry curvatures of the FM HC lattice given by Eq.\,(\ref{bfhc}).
Using the same parameters as in Figs.\,7 and 8, we plot $\Omega_{1z}(\vk )$ in Fig.\,9.  

\begin{figure}
\begin{center}
\includegraphics[width=9cm]{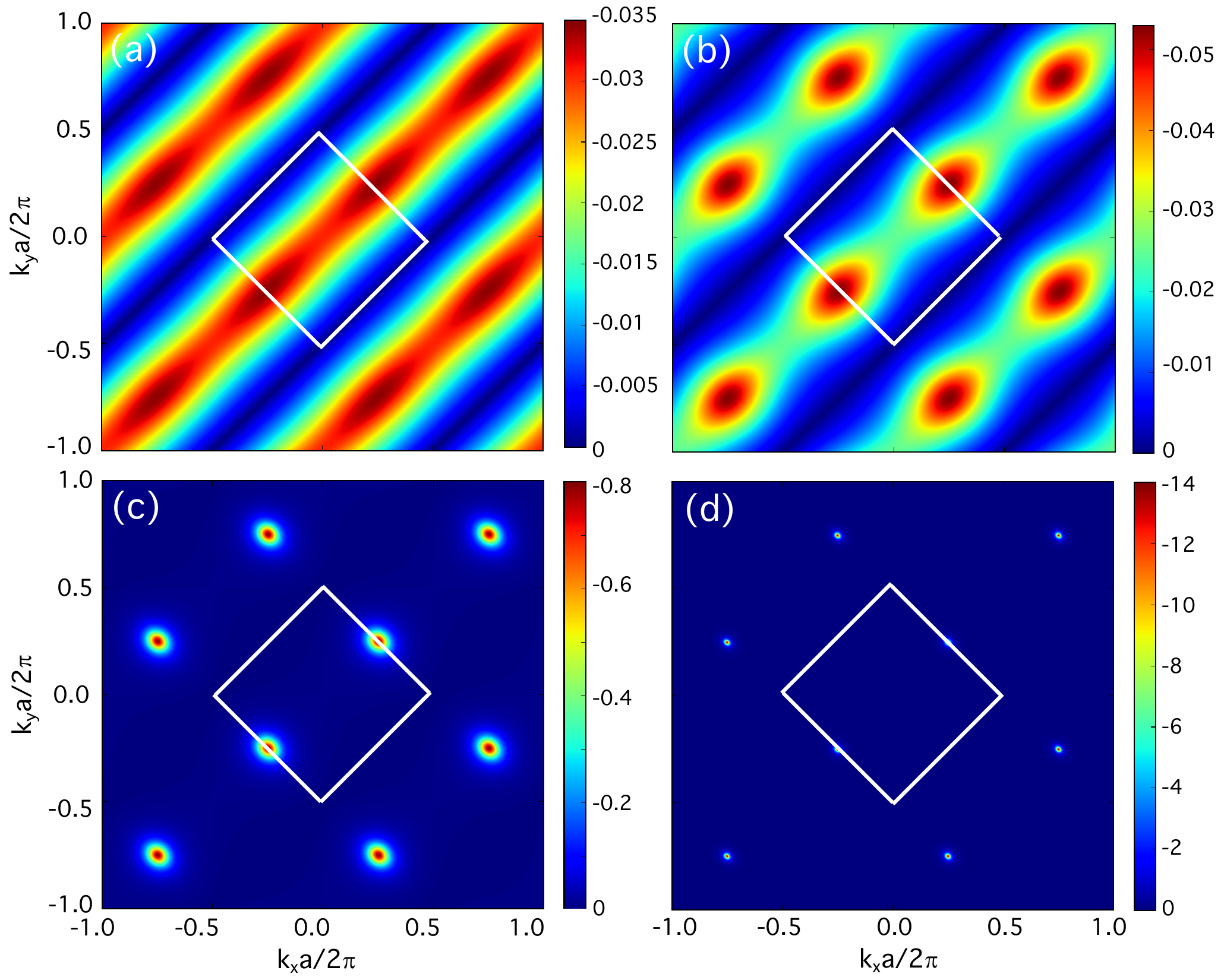}
\end{center}
\caption{The Berry curvature $\Omega_{1z}(\vk )$ of the FM ZZ lattice evaluated using $d=-0.4$ and $r$ = (a) 50, (b) 8, (c) 1.5, and (d) 1.1.  The first BZ
of the magnetic unit cell is sketched by the solid lines.}
\label{Fig9}
\end{figure}

Comparing Figs.\,6 and 9 reveals that the Berry curvature 
vanishes at the upper left and lower right boundaries of the BZ with $k_x-k_y=\pm \pi/a$,
where the magnon gap $\delta (\vk )$ also vanishes.  For any $r$, the DM interaction does not affect
the gap at $k_x=k_y =\pm 0.5\pi/a$, where $\tau_{\vk }=0$ and $\delta (\vk ) = 2\vert \Psi_{\vk }\vert =2\vert 1-r\vert /(1+r)$.  So at $r= 1$, the gap between the bands closes
at those two points.  Close to $r=1$, strong peaks in the Berry curvature are found at those same $\vk $ points in Fig.\,9.   However, 
the Berry curvature disappears when the exchange becomes homogeneous as $r\rightarrow 1$.  
 

Evaluating the Chern number $C_n$ for the FM ZZ model by integrating $\Omega_{nz}(\vk )$ over all $\vk $ within the first BZ zone, we obtain the surprising result 
that $C_n$ are not well defined.  
Recall that the Chern numbers for the FM HC lattice are $\pm 1$ for all $d$.  The Chern numbers for the FM ZZ model are not well defined due to 
the degeneracy of the magnon bands at the upper right and lower left boundaries of the BZ.  


\section{Anisotropic FM Zig-zag Lattice}

As mentioned earlier, the degeneracy of the magnon bands along the BZ boundaries 
can be lifted by allowing the exchanges $J_{nx}$ along the $x$ axis to be different than the exchanges $J_{ny}$ along the $y$ axis.
Earlier results must then be modified by defining $J_t=J_{1x}+J_{1y}+J_{2x}+J_{2y}$, $d=-4D/J_t$,
\begin{equation}
\Psi_{\vk }=\frac{J_{1x}e^{-ik_xa}+J_{iy}e^{-ik_ya}+J_{2x}e^{ik_xa}+J_{2y}e^{ik_ya}}{J_t},
\end{equation}
\begin{equation}
\hbar \omega_1(\vk)=J_tS ( 1  - \mu_{\vk } + \kappa )
\label{zzfm1r},
\end{equation}
 \begin{equation}
\hbar \omega_2(\vk)=J_tS ( 1  + \mu_{\vk } + \kappa ),
\label{zzfm2r}
\end{equation}
\begin{equation}
r=\frac{J_{2x}+J_{2y}}{J_{1x}+J_{1y}} >1,
\end{equation}
and $\delta (\vk )=\hbar \Delta \omega(\vk)/J_tS$. 

The anisotropic FM ZZ model contains three regimes depending on the relative values of
$\Delta J_1 = \vert J_{1y}-J_{1x}\vert $ and $\Delta J_2 = \vert J_{2y}-J_{2x}\vert $, with the constraint that $r >1$.
For the case $\Delta J_1=\Delta J_2 =0$ considered in the previous section, 
the Chern number is undefined and the gap between the bands vanishes even when $d\ne 0$.   The Chern number number 
remains undefined when $\Delta J_1=\Delta J_2 > 0$, although the gap between the bands is then nonzero due to the exchange anisotropy
in the $x$ and $y$ directions.
For $\Delta J_1 > \Delta J_2\ge 0$, the Chern numbers $C_n$ for the lower and upper bands are $-1$ and $+1$ when $d<0$ and reversed when
$d>0$.  When $\Delta J_2 > \Delta J_1\ge 0$, $C_n=0$.  Consequently, the anisotropic FM ZZ model bears some similarity to the 
Su-Schrieffer-Heeger model \cite{Su1979}.

For simplicity, we now consider the case where $\Delta J_1\ne 0$ but $\Delta J_2=0$.
Using $J_{1y}/J_{1x}=1.5$, the normalized gap $\delta (\vk )$ is plotted versus $\vk $ in Fig.\,10.   
The minimum value of $\delta (\vk )$ increases from $2.9 \times 10^{-3}$ to $2.5 \times 10^{-2}$ as $\vert d\vert $ increases from 0 to 0.4.  
The short-dash curve in Fig.\,8 for $r=1.5$ uses $J_{1y}/J_{1x}=1.1$, indicating a redistribution of OAM to values of $k$ near its peak. 

The Berry curvatures of the anisotropic FM ZZ model with $J_{1y}/J_{1x}=1.5$, $\Delta J_2=0$, and $d=-0.4$ are plotted in Fig.\,11 
for four different values of $r$.  Notice that the range of Berry curvatures now extends over
both positive and negative values with the upper negative bounds for $d=-0.4$ exceeding the range for $J_{1y}/J_{1x}=1$ in Fig.\,9.   
These negative bounds for the Berry curvature can be found on the upper left
and lower right boundaries of the BZ where the magnon modes were degenerate and the Berry curvature vanished for $J_{1y}/J_{1x}=1$.

In Appendix D, we construct a ribbon with edges along the zig zags to 
demonstrate that non-trivial topological edge modes are associated with the model when the Chern numbers are $\pm 1$.  No such edge modes appear
when the Chern numbers are undefined or 0.

\begin{figure}
\begin{center}
\includegraphics[width=9cm]{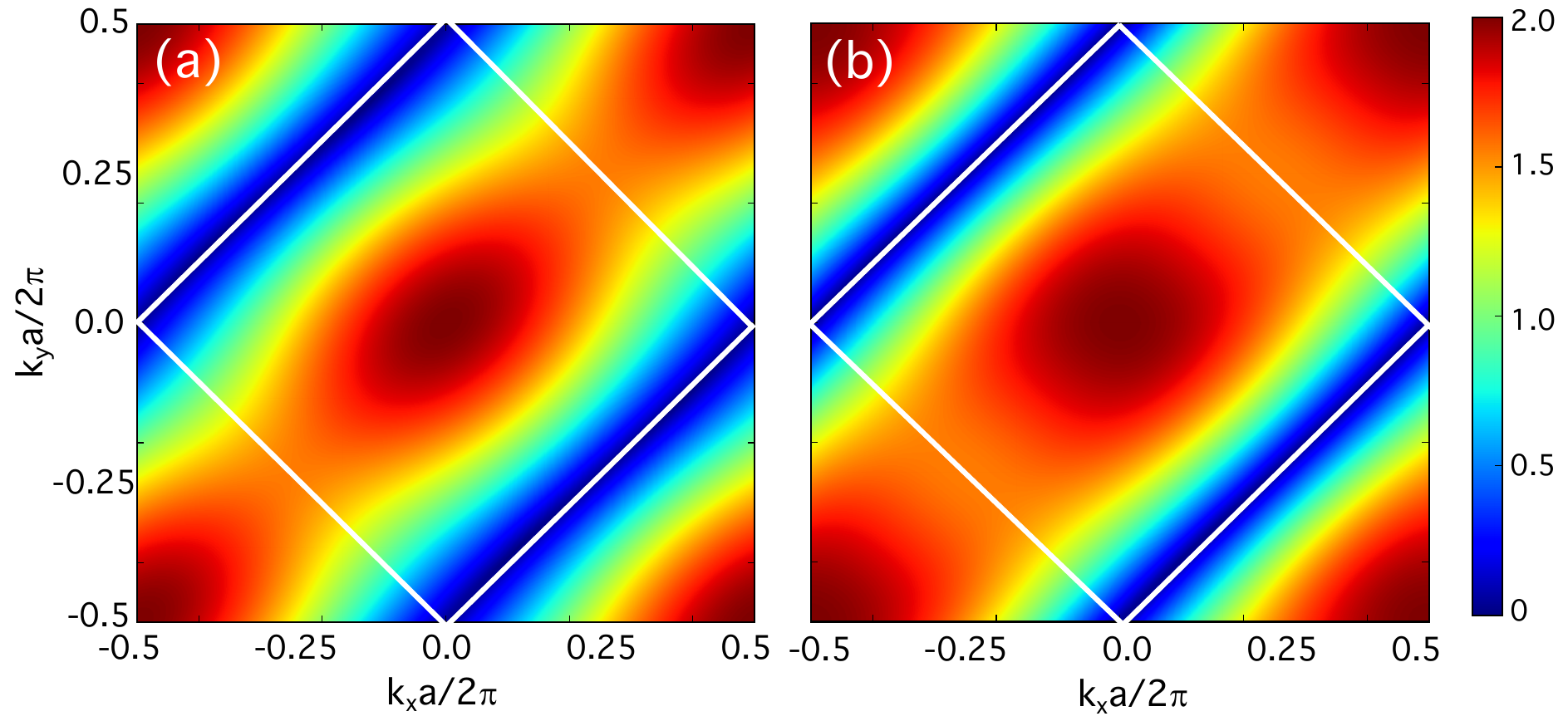}
\end{center}
\caption{The normalized gap $\delta (\vk )=2\mu_{\vk }$ between bands of the FM ZZ lattice evaluated using $J_{1y}/J_{1x}=1.5$, $\Delta J_2=0$, $r=8$, and (a) $d=0$ or (b) $-0.4$.
The first BZ of the magnetic unit cell is sketched by the solid lines.}
\label{Fig10}
\end{figure}

\begin{figure}
\begin{center}
\includegraphics[width=9cm]{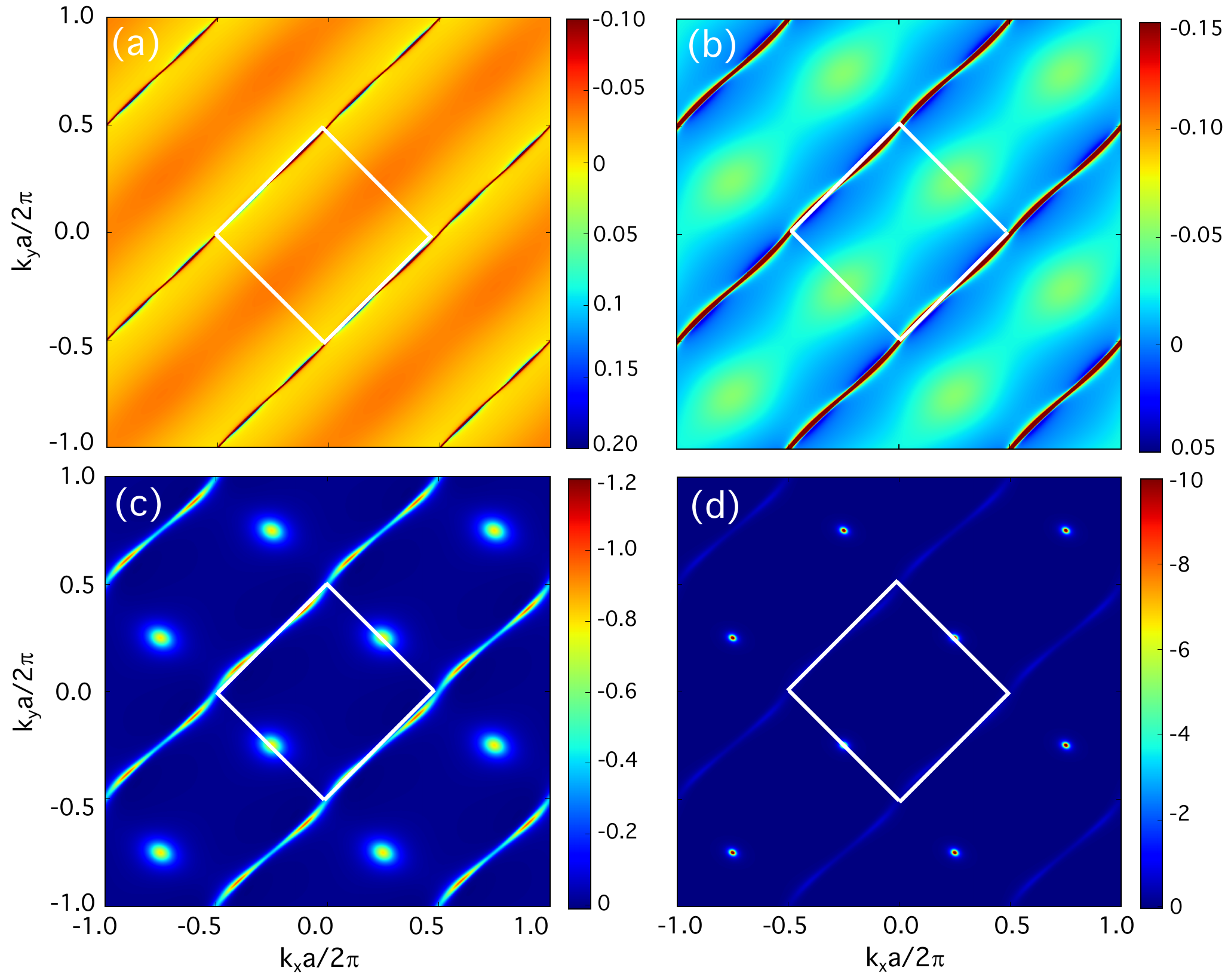}
\end{center}
\caption{The Berry curvature $\Omega_{1z}(\vk )$ of the FM ZZ lattice evaluated using $d=-0.4$, $J_{1y}/J_{1x}=1.5$, $\Delta J_2=0$, and $r$ = (a) 50, (b) 8, (c) 1.5, and (d) 1.1.  The first BZ
of the magnetic unit cell is sketched by the solid lines.}
\label{Fig11}
\end{figure}

\section{Discussion and Conclusion}

The magnon Hall effect was first predicted for a FM Kagom\'e lattice \cite{Katsura2010} with DM interactions due
to broken inversion symmetry.   The subsequent observation and theory of the 
magnon Hall effect was performed for FM pyrochlore systems \cite{Onose2010, Ideue12}.  Nonzero Berry curvatures and Chern numbers were 
also suggested for the FM star lattice \cite{Owerre17}, which has similarities to both Kagom\'e and HC lattices.
Earlier work predicted \cite{Cheng16, Owerre16, Fishman23b} that OAM can be observed
in FM HC lattices.  Our current paper 
predicts that OAM can also be observed in FM ZZ lattices with distinct exchange interactions $0 < J_1 < J_2$ and $D\ne 0$.
As shown in Appendices B and C, 
OAM is not observable in AF HC and ZZ geometries even when $D\ne 0$. 

The finite Berry curvatures in both FM HC and ZZ models implies that their thermal conductivities are nonzero.  The magnon Hall effect 
\cite{Shindou13} is evaluated in terms of the Berry curvature using Eq.\,(\ref{kxy}), where
\begin{equation}
c_2(\rho)=(1+\rho )\Bigl(\log \frac{1+\rho}{\rho } \Bigr)^2-(\log \rho )^2 -2 {\rm Li}_2 (-\rho)
\end{equation}
and ${\rm Li}_2(z)$ is the dilogarithmic function.
To account for the different scaling of the magnon energies
$\epsilon_n(\vk )=\hbar \omega_n (\vk )$ for the two FM models, we define $\tilde T=\kB T/3JS$ and 
$\tilde \kappa^{xy }=\hbar \kappa^{xy }/3\kB JS$ (HC) or $\tilde T=\kB T/2(J_1+J_2)S$ and
$\tilde \kappa^{xy }=\hbar \kappa^{xy }/2\kB (J_1+J_2)S$ (ZZ)
with the dimensionless thermal conductivity given by
\begin{equation}
\tilde \kappa^{xy}({\tilde T})=-\frac{\tilde T}{2\pi }\sum_n \int_{BZ} d^2k\, c_2\bigl(\rho(\epsilon_n(\vk ))\bigr)\,\Omega_{nz}(\vk )
\end{equation}
for both models.

To examine the effect of the ratio $r=J_2/J_1 >1$ for the FM ZZ model, 
we plot $\tilde \kappa^{xy}(\tilde T)$ versus $r$ for several values of $d$ in Fig.\,12, which sets $\tilde T=0.3$, and $\kappa=1.5$.
As expected $\tilde \kappa^{xy}(\tilde T)\rightarrow 0$ as $r\rightarrow 1$.  More unexpectedly, $\tilde \kappa^{xy}(\tilde T)$ reaches a plateau at about $r \approx 6$,
which implies that the magnon Hall effect will be most easily observed in materials with large $r$.

\begin{figure}
\begin{center}
\includegraphics[width=7cm]{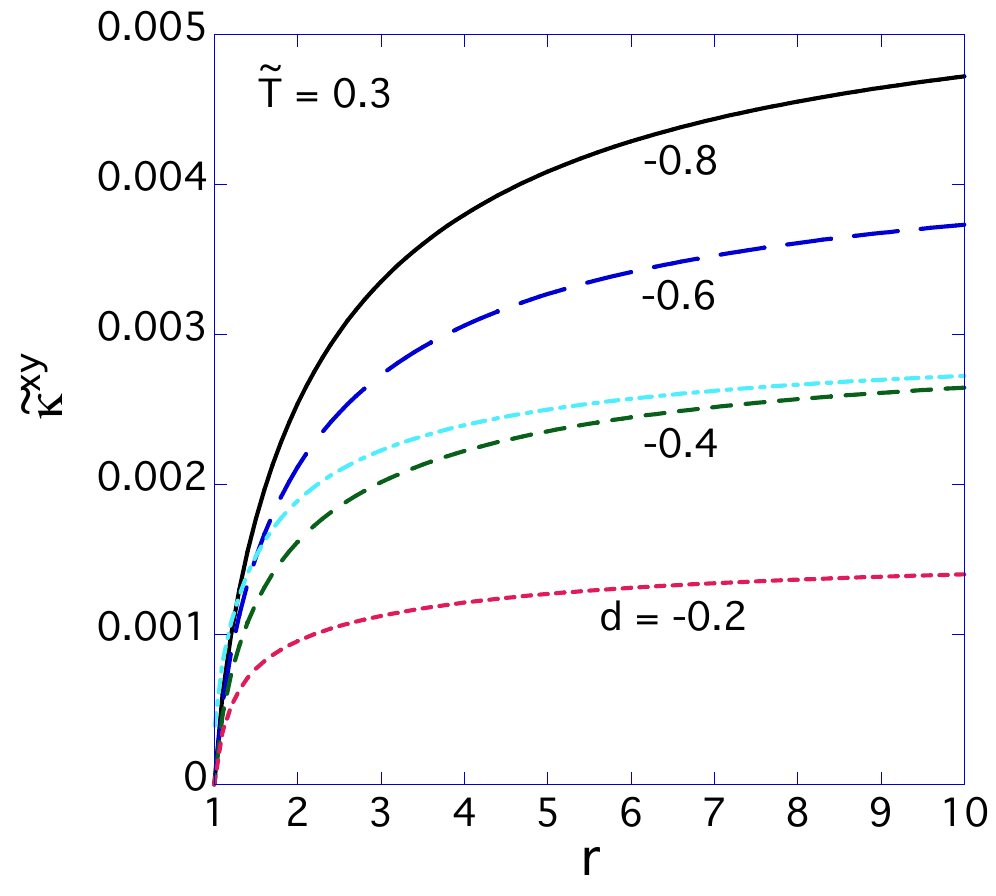}
\end{center}
\caption{The normalized thermal conductivity $\tilde \kappa^{xy}(\tilde T)$ versus $r$ for several values of $d$ for the FM ZZ model with temperature 
$\tilde T=0.3$ and anisotropy $\kappa =1.5$.  The dot-dash curve for $d=-0.4$ takes $J_{1y}/J_{1x}=1.5$.}
\label{Fig12}
\end{figure}

The magnon Hall effect has been observed in experimental realizations of the pyrochlore \cite{Onose2010, Ideue12} and kagome \cite{Hirschberger15b} lattices.
Unfortunately, it has not yet been observed in FM HC lattices like CrBr$_3$ \cite{Cai21} and CrIr$_3$ \cite{Chen18}.  
Nevertheless, we compare theoretical values of $\tilde \kappa^{xy}(\tilde T)$ for the FM HC and ZZ models in Fig.\,13.  Taking $r=10$
for the FM ZZ model and setting $\tilde T\approx 0.6$, we find that 
$\tilde \kappa^{xy}(\tilde T)$ is about four times larger for the HC model than for the ZZ model.  The suppression of the FM state with temperature can always be
alleviated by increasing the anisotropy $K$.

In order to estimate the effect of 
the partially gapped magnons on the ZZ model, we plot the thermal conductivities with 
$J_{1y}/J_{1x}=1.5$ in Figs.\,12 and 13.  Notice that the magnon band gap produced by $J_{1y}/J_{1x}=1.5$ enhances $\tilde \kappa^{xy}(\tilde T)$ 
only slightly, with the largest change at small $r$ in Fig.\,12.   
So the disappearance of the magnon band gap and the absence of a well-defined Chern number when $J_{1y}/J_{1x}=1$ does
not significantly depress the magnon thermal conductivity at large $r$.

\begin{figure}
\begin{center}
\includegraphics[width=7cm]{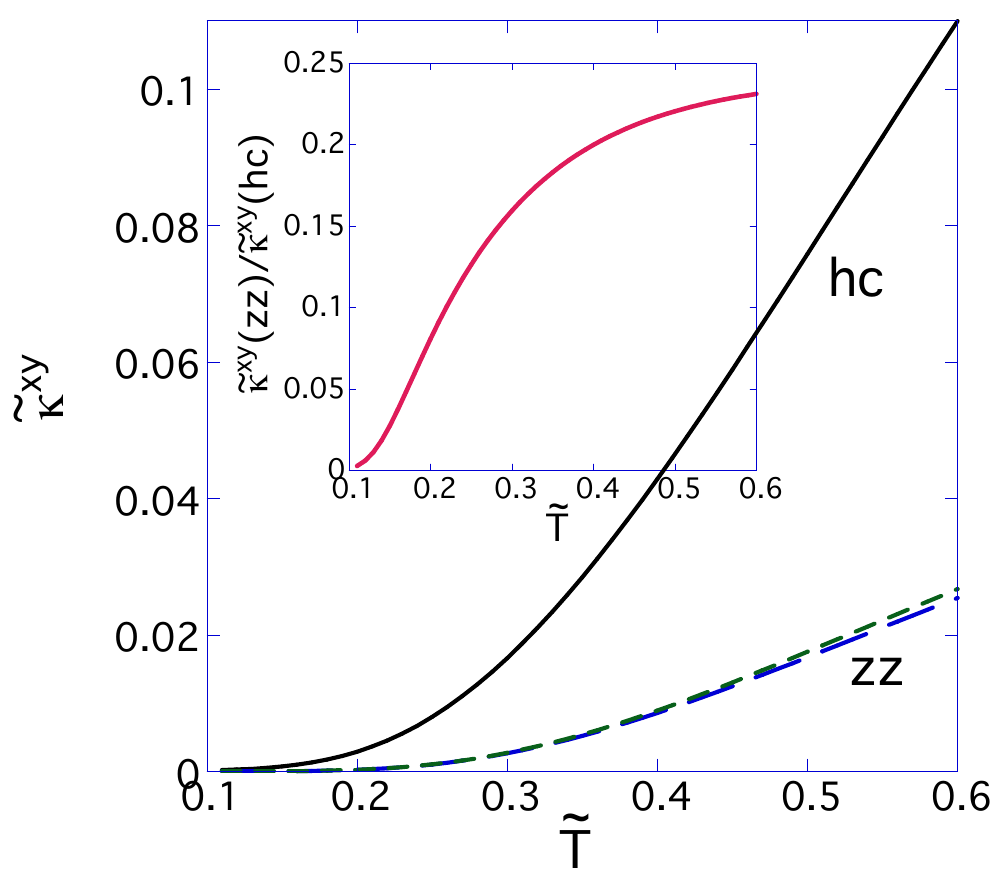}
\end{center}
\caption{The normalized thermal conductivity $\tilde \kappa^{xy}(\tilde T)$ of the FM HC (solid) and ZZ (long dash) models versus $\tilde T$.
Both models take $d=-0.4$ and $\kappa =1.5$ while the FM ZZ model also uses $r=10$ and the upper (small dash) ZZ curve 
sets $J_{1y}/J_{1x}=1.5$.  The inset plots the ratio of the normalized thermal conductivities of the ZZ and HC lattices versus 
$\tilde T$ with $J_{1y}/J_{1x}=1$.}
\label{Fig13}
\end{figure}

Due to the low contrast between different FM exchange couplings, it is difficult to identify materials described by the FM ZZ geometry.
Nevertheless, several cases of FM ZZ chains coupled by FM exchange interactions have been discovered:  
spin-1/2 Heisenberg Vanadium chains in CdVO$_3$ \cite{Onoda99, Tsirlin11}, spin-3/2 Chromium chains in LaCrOS$_2$ \cite{Takano02}, 
and spin-3.4/2 Manganese chains in La$_3$MnAs$_5$ \cite{Duan22}.   For CdVO$_3$ \cite{Onoda99, Tsirlin11}, the intrachain coupling 
$J_2\approx 90$\,K is significantly stronger than the interchain coupling
$J_1 \approx 18$\,K so that $r\approx 5$.   For La$_3$MnAs$_5$ \cite{Duan22}, $r \approx 7.6$.  
The exchange ratio $r$ is also believed to be large in LaCrOS$_2$ \cite{Takano02}.
As predicted by Fig.\,12, observation of the magnon Hall effect
sensitively depends on the ratio $r=J_2/J_1$.  With such large values of $r$, any of the materials mentioned above will be good candidates to search
for the magnon Hall effect in FM ZZ geometries.

Edge currents produced by the Berry curvature \cite{Mong11,Mat11a, Mat11b} and closely connected with the thermal conductivity and Chern number \cite{Zhang13, Mook14a}
are only topologically protected in systems containing 
a gap between the magnon bands, i.e. in magnetic insulators. 
Therefore, the edge currents in FM ZZ lattices with $J_{1y}/J_{1x}=1$ are not topologically protected and will decay with time due to the degenerate 
$\vk $-space regions along the $k_x-k_y =\pm \pi /a$ BZ boundaries 
where the magnon bands overlap.  The topological protection afforded by the symmetry breaking $J_{1y}/J_{1x}\ne 1$ of the exchange interaction $J_1$ 
may then depend on the size of the resulting magnon band gap compared to the temperature and the magnon interactions \cite{Mook21}.

Direct observation of the angular-averaged magnon OAM $F_n(k)$ may be possible in both HC and ZZ lattices by coupling magnons to other particles and quasiparticles.
While the dynamics of $F_n(k)$ remains unexplored, we speculate that the magnon OAM may appear in inelastic neutron scattering experiments through
the angular-averaged $S(\vk ,\omega )$, i.e. the powder-averaged $S(k,\omega )$, when $\omega $ crosses a magnon band and $k$ crosses a peak in 
$F_n(k)$.  Alternatively, it may be possible to couple magnons to chiral phonons in either FM honeycomb or zig-zag materials.  
For example, chiral phonons and their quasiparticle couplings have been observed in the honeycomb structure WSe$_2$ \cite{Zhu18} while the 
FM zig-zag material La$_3$MnAs$_5$ has a 63 screw axis that supports phonons with OAM \cite{Juneja22}.
It may also be possible to couple the magnons to a high energy electron beam after the 
electrons are separated into orbital components by a grating \cite{McMorran17}.  A STEM can then be used to probe the coupling between the electron 
and magnon OAM when the transverse linear electron momentum matches the linear magnon momentum.

To conclude, we have studied a new class of materials associated with FM ZZ geometries where the effects of OAM are observable.
Formally, results for the OAM and Berry curvature for this geometry are quite similar to well-known results for the FM HC lattice.
While the magnon bands are not completely gapped and the Chern numbers are not well defined for $J_{1y}/J_{1x}=1$, those deficits 
do not significantly impact the magnon thermal conductivity $\kappa^{xy}(T)$ in ZZ lattices.  
Indeed, opening a magnon band gap and producing well-defined Chern numbers by setting $J_{1y}/J_{1x}\ne 1$
only modestly enhances $\kappa^{xy}(T)$.  Although only an infinitesimal difference $J_{1y}/J_{1x}-1\ne 0$ is required to create a magnon gap,
FM ZZ lattice materials with $J_{1y}/J_{1x}=1$ are not topological insulators.  Consequently, the usefulness of these materials for specific 
applications may depend on the lifetime of the edge modes.
Nonetheless, we are hopeful that future experiments on some of the materials discussed above
will demonstrate that the effects of OAM may be observed in systems that are neither topological nor magnetic insulators. 

Conversations with Shulei Zhang are gratefully acknowledged.  
Research by R.F., L.L. and T.B. 
sponsored by the U.S. Department of Energy, 
Office of Science, Basic Energy Sciences, Materials Sciences and Engineering Division.
Theoretical contributions by J.V. were supported by the U.S. Department of Energy, Office of Science, 
National Quantum Information Science Research Centers, Quantum Science Center.
The data that support the findings of this study are available from the authors
upon reasonable request.

\appendix

\section{Tiling OAM}

For the FM ZZ lattice, the smoothness of the even part of the OAM at the zone boundaries is ensured if 
$O_{1,2}^{({\rm even})} (k_x,k_y )=O_{1,2}^{({\rm even})} (k_y,k_x)$, as seen numerically in Figs.\,3 and 7. 
For this case, a point just outside the BZ can be mapped into the corresponding point on the opposite side of the boundary just inside the BZ by 
using
\begin{equation} 
O_{1,2}^{({\rm even})} (\vk)=O_{1,2}^{({\rm even})} (\vk+\vG)
\end{equation}
and  
\begin{equation}
O_{1,2}^{({\rm even})} (k_x,k_y)=O_{1,2}^{({\rm even})}  (-k_y,-k_x),
\end{equation}
where $\vG $ is a reciprocal lattice vector and we have used the even property of $O_{1,2}^{({\rm even})}(\vk )$. 
Thus, in approaching the BZ edge, the even OAM of point $\vk $ and its mirror across the BZ boundary are equivalent, all the way to the limit of the boundary itself. 
To show that $O_{1,2}^{({\rm even})} (k_x,k_y )=O_{1,2}^{({\rm even})} (k_y,k_x)$ for the FM ZZ square lattice, 
we see that $O_{1,2}^{({\rm even})}(\vk )$  in Eq.\,(48) is composed of functions that are symmetric or antisymmetric with respect to switching components: 
$\Psi_{(k_x,k_y)}=\Psi_{(k_y,k_x)}$, 
$\tau_{(k_x,k_y)}=-\tau_{(k_y,k_x)}$, $\mu_{(k_x,k_y)}=\mu_{(k_y,k_x)}$, and $\hat l_z(k_x,k_y)= -\hat l_z(k_y,k_x)$. Plugging these into Eq.\,(48), 
the negative signs from the antisymmetric terms cancel and $O_{1,2}^{({\rm even})}(k_x,k_y )=O_{1,2}^{({\rm even})} (k_y,k_x)$.

\section{AF Honeycomb Lattice}

This appendix considers the HC lattice sketched in Fig.\,1(b) with AF exchange  
$J<0$ between alternating up and down spins.  We then find
\begin{equation}
\underline{L}(\vk ) =
-\frac{3JS}{2} \left(
\begin{array}{cccc}
A^+_{\vk } & 0 & 0 & -\Gamma_{\vk}^* \\
0 & A^+_{\vk } &  -\Gamma_{\vk} & 0\\
0 & -\Gamma_{\vk}^* & A^-_{\vk } & 0\\
-\Gamma_{\vk} & 0 &0 & A^-_{\vk } \\
\end{array} \right),
\end{equation}
where $A^{\pm }_{\vk }=1 \pm d\, \Theta_{\vk } + \kappa $ as in the FM HC lattice but with $\kappa = 2K/3\vert J\vert $.
It is then easy to show that solutions for the eigenfunctions $X^{-1}_{rn}(\vk )$ are independent of $d$.
The doubly degenerate magnon energies  
\begin{equation}
\hbar \omega_{1,2}(\vk )=3\vert J\vert S\sqrt{(1+\kappa )^2-\vert \Gamma_{\vk }\vert^2} + d\,\Theta_{\vk },
\end{equation}
are simply shifted by the DM interaction.  As expected, $O_n(\vk )$ is an odd function of $\vk $ for any gauge.
Therefore, the AF HC lattice does not support an observable OAM and $F_n(k)=0$.  However, the AF HC lattice does 
support the magnon Nernst effect with a net spin current \cite{Kim16, Shiomi17}.

\section{AF Zig-zag Lattice}

This appendix treats the AF ZZ lattice with AF coupling $J_1 < 0$ between chains and FM coupling $J_2 > 0$ within chains.  
As seen in Fig.\,1(d), the magnetic unit cell contains 4 spins so the $\uL (\vk )$ matrix is 8 dimensional.
Fortunately, we can write
\begin{eqnarray}
&&H_2={\sum_\vk }' {\bf v}_{\vk}^{\dagger }\cdot \underline{{\it L}}(\vk )\cdot {\bf v}_{\vk }\nonumber \\
&&={\sum_\vk }' \Bigl\{ {\bf v}_{1\vk}^{\dagger }\cdot \uL_1(\vk )\cdot {\bf v}_{1\vk } + {\bf v}_{2\vk}^{\dagger }\cdot \uL_2(\vk )\cdot {\bf v}_{2\vk }\Bigr\},
\label{defL12}
\end{eqnarray}
where
\begin{eqnarray}
{\bf v}_{1\vk } &=&(a_{\vk }^{(1)},a_{\vk }^{(2)},a_{-\vk}^{(3)\dagger },a_{-\vk }^{(4)\dagger }),\\
{\bf v}_{2\vk } &=&(a_{\vk }^{(3)},a_{\vk }^{(4)},a_{-\vk}^{(1)\dagger },a_{-\vk }^{(2)\dagger }),
\end{eqnarray}
\vskip .3cm
\begin{equation}
\uL_1(\vk ) =
S(J_2-J_1) \left(
\begin{array}{cccc}
A^-_{\vk } & -\gamma_2\xik  & 0 & \gamma_1\xik^* \\
-\gamma_2\xik^* & A^+_{\vk } &  \gamma_1\xik  & 0\\
0 & \gamma_1\xik^* & A^-_{\vk } & -\gamma_2\xik \\
\gamma_1 \xik  & 0 &-\gamma_2\xik^* & A^+_{\vk } \\
\end{array} \right),
\end{equation}
\vskip.5cm
\begin{equation}
\uL_2(\vk ) =
S(J_2-J_1) \left(
\begin{array}{cccc}
A^+_{\vk } & -\gamma_2\xik  & 0 & \gamma_1\xik^* \\
-\gamma_2\xik^* & A^-_{\vk } &  \gamma_1\xik  & 0\\
0 & \gamma_1\xik^* & A^+_{\vk } & -\gamma_2\xik \\
\gamma_1 \xik  & 0 &-\gamma_2\xik^* & A^-_{\vk } \\
\end{array} \right),
\end{equation}
with $A^{\pm }_{\vk }=1\pm d\,\tau_{\vk }$, $d=2D/(J_2-J_1)$, and $\gamma_n=J_n/2(J_2-J_1)$.
The only difference between $\uL_1(\vk )$ and $\uL_2(\vk )$ is that $D$ changes sign.
It is straightforward to show that the symmetry relations of Eqs.\,(\ref{up}) and (\ref{uq}) are
satisfied by the full matrix $\uL (\vk)$.

The AF ZZ model then contains 4 magnon bands, which are doubly degenerate with energies
\begin{eqnarray}
&&\hbar \omega_{1,3}(\vk )= 2(J_2-J_1)S\biggl\{ 1 - (\gamma_1^2-\gamma_2^2)\vert \xik\vert^2 +16(d\, \tau_{\vk })^2\nonumber \\
&&\pm  \sqrt{ \gamma_2^2 \Bigl( \gamma_1^2 (\xik^2 -\xik^{*2})^2+4\vert \xik\vert^2\Bigr) + 4(d \,\tau_{\vk})^2 } \biggr\}^{1/2},
\end{eqnarray}
$\omega_2(\vk )=\omega_1(\vk)$, and $\omega_4(\vk )=\omega_3(\vk )$.
Since the magnon energy does not depend on the sign of $d$, lower bands 1 and 2 and upper bands 3 and 4 from $\uL_1(\vk )$ and $\uL_2(\vk )$ are degenerate.

Additional exchange interactions do not affect the structure of the $\uL_n(\vk )$ matrices.  For example, an exchange interaction $J_3$ between spin pairs
$\{1,3\}$ and $\{2,4\}$ along the $(1,1)$ diagonal does not couple the $\uL_1(\vk )$ and $\uL_2(\vk )$ matrices.  Nor do any other complex set of exchange interactions or anisotropies.
So the partition of $\uL (\vk )$ into two $4\times 4$ matrices remains unaltered.

Numerical calculation of $F_n(k)$ reveals that $F_1(k)=-F_2(k)$ and $F_3(k)=-F_4(k)$ so that the contribution of the two $\uL_n(\vk )$ matrices with opposite $D$ cancel.
Hence, there is no {\it net} observable OAM from the two lower or two upper magnon bands.  
Similarly, we find that the Berry curvatures of bands 1 and 2 cancel, as do the Berry curvatures of bands 3 and 4.

\section{Edge modes for the Anisotropic FM Zig-zag lattice}

The anisotropic FM ZZ model has Chern numbers $C_n=\pm 1$ when $\Delta J_1 > \Delta J_2$, $C_n=0$ when $\Delta J_2 > \Delta J_1$, and $C_n$
undefined when $\Delta J_1=\Delta J_2$, all with the understanding that $r>1$.  In this Appendix, we show that a ribbon cut along the zig zags
contains topological edge modes only when $\Delta J_1 > \Delta J_2$.  

\begin{figure}
\begin{center}
\includegraphics[width=7cm]{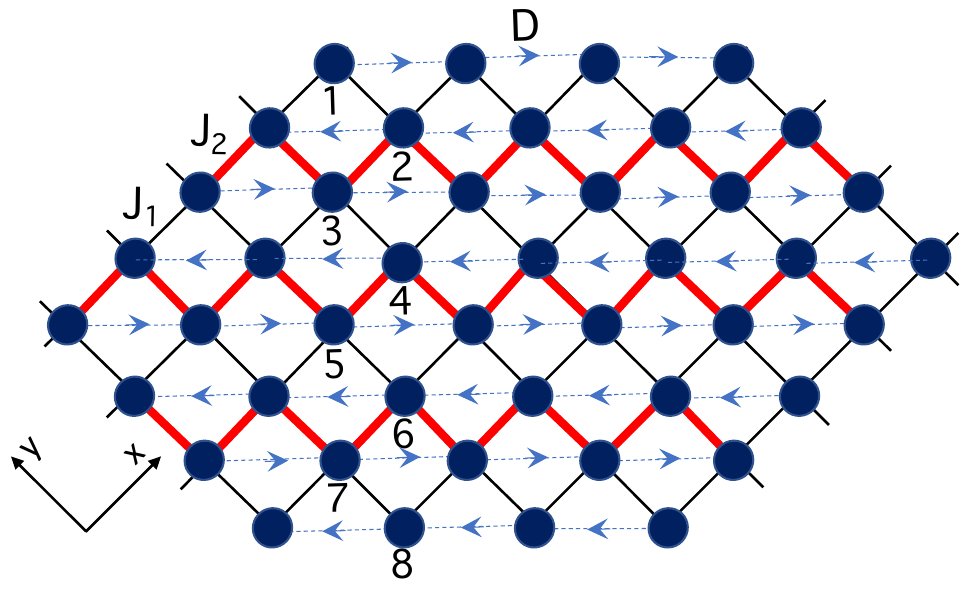}
\end{center}
\caption{A ribbon with width $M=8$ containing weaker bonds $J_{1n}$ at the outer edges.  The termination is only at the top and bottom, the ribbon propagates to infinity in the other directions.}
\label{Fig14}
\end{figure}

There are three ways to construct a ribbon with width $M$:  with the outer bonds given by the stronger exchange interactions $J_{2n}$ ($M$ even), 
given by the weaker exchange interactions $J_{1n}$ ($M$ even), or with $J_{2n}$ on one side and $J_{1n}$ on the other ($M$ odd).  In Fig.\,14, we have sketched a ribbon 
with $M=8$ and the outer bonds given by $J_{1n}$.  For this kind of ribbon, the $2M \times 2M$ $\underline L(\vk )$ matrix is constructed by taking
\begin{eqnarray}
&&L_{11}(\vk)=L_{MM}(\vk)=\frac{S}{2}(J_{1x}+J_{1y}) \nonumber \\
&& \mp \frac{SJ_t}{2}d\,\sin(k_ya-k_xa),
\end{eqnarray} 
\begin{eqnarray}
&&L_{rr}(\vk ) = \frac{SJ_t}{2} \nonumber \\
&+&\frac{SJ_t}{2} (-1)^r d\,\sin(k_ya-k_xa),\,\,\, r\ne1,M,
\end{eqnarray} 
\begin{equation}
L_{r,r+1}(\vk )=-\frac{S}{2}(J_{nx}e^{-ik_xa}+J_{ny}e^{-ik_ya}),
\end{equation}
with $n=1$ for $r$ odd and $n=2$ for $r$ even.  We also have
$L_{r+1,r}(\vk )=L_{r,r+1}(\vk )^{\star}$ 
for $r$ both odd or even.  Note that in all matrix elements $L_{rs}(\vk)$, $r$ and $s$ are assumed to be less than or equal to $M$.  
Then the matrix elements $L_{r+M,s+M}(\vk )$ are determined by using Eq.\,(\ref{up}).

The differences between the outer and inner bonds are that the outer bonds have different diagonal terms $L_{11}(\vk )$ and $L_{MM}(\vk )$
because they only couple to interior sites through the weaker exchange interactions $J_{1n}$.  Also sites 1 couple only downwards to sites 2 and sites $M$
only couple upwards to sites $M-1$ through those weaker bonds.  Notice that the DM interaction alternates directions from the top of the ribbon to the 
bottom but that its strength remains the same.

\begin{figure}
\begin{center}
\includegraphics[width=8.3cm]{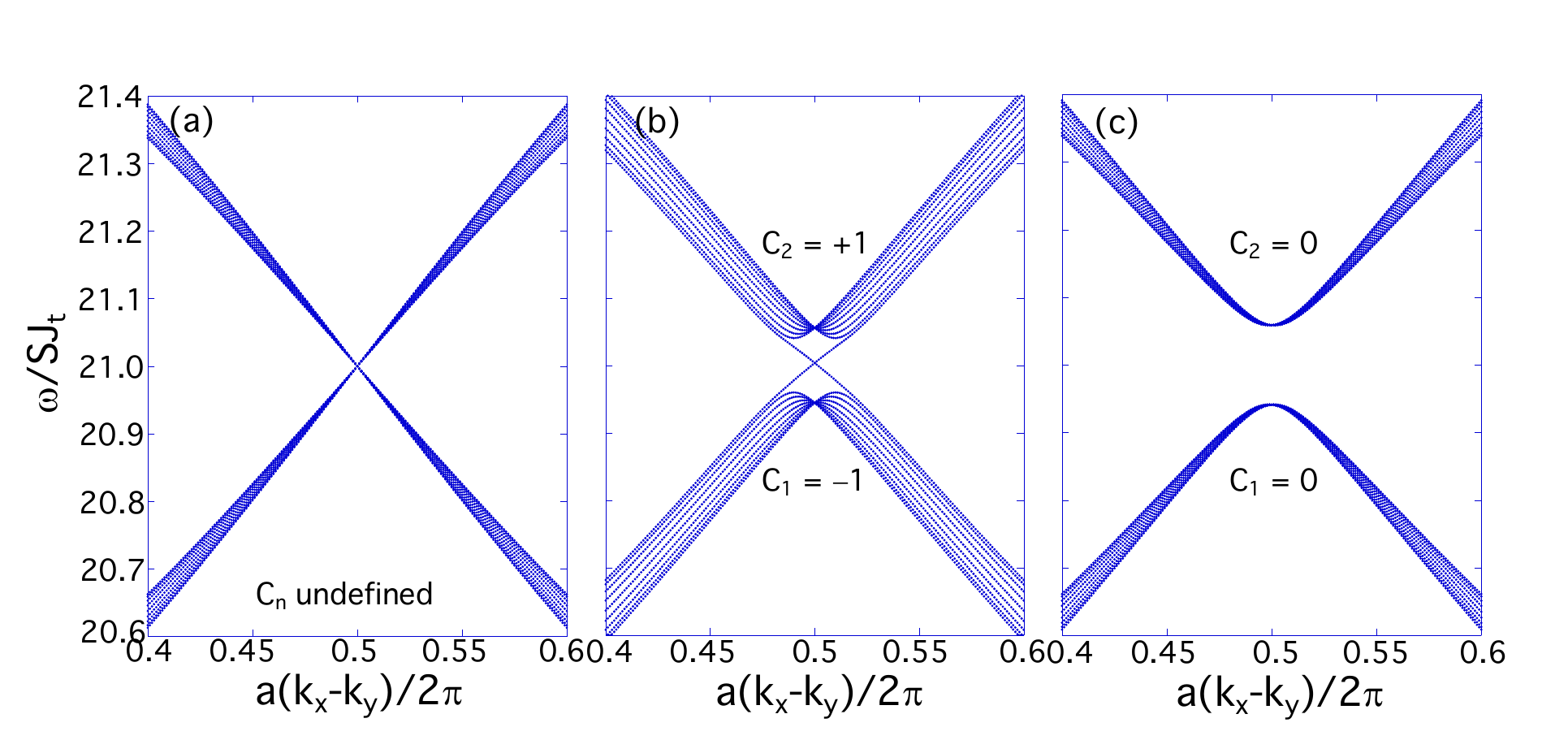}
\end{center}
\caption{The frequencies $\omega /SJ_t$ for a ribbon with width $M=20$ containing weaker bonds $J_{1n}$ at the outer edges and (a) $J_{1x}=J_{1y}=1$, $J_{2x}=J_{2y}=8$,
(b) $J_{1x}=0.5$, $J_{1y}=1.5$, $J_{2x}=J_{2y}=8$, or (c) $J_{1x}=J_{1y}=1$, $J_{2x}=7$, $J_{2y}=8$, all with $\kappa =10$ and $d=-0.4$.}
\label{Fig15}
\end{figure}

Results for the modes of this type of ribbon are shown in Fig.\,15.   In Fig.\,15(a) with $\Delta J_1=\Delta J_2=0$, the Chern number $C_n$ is undefined and there is no
gap due to the absence of $xy$ anisotropy.  In Fig.\,15(b) with $\Delta J_1 > \Delta J_2=0$, a band gap appears due to the $xy$ anisotropy in $J_{1n}$.  Topological edge modes are associated 
with the Chern numbers $C_1=-1$ and $C_2=+1$.
In Fig.\,15(c) with $\Delta J_2 > \Delta J_1=0$, a band gap once again appears due to the $xy$ anisotropy (now in $J_{2n}$) but the edge modes are absent and $C_1=C_2=0$.   
Hence, there is a clear connection between our earlier results for the Chern numbers and the appearance of topological edge modes on a ribbon.
This is called \cite{Mong11} the bulk-edge or bulk-boundary correspondence.

\vfill\eject

\end{document}